%% file: main_MS.tex
\documentclass[A5,11pt]{article}
\usepackage[margin=1in]{geometry}

\usepackage{natbib}
 \bibpunct[, ]{(}{)}{,}{a}{}{,}%
 \def\bibfont{\small}%
\makeatletter
\renewcommand\@afterheading{%
  \@nobreaktrue
  \everypar{%
    \if@nobreak
      \@nobreakfalse
      \clubpenalty 1
      \if@afterindent \else
        {\setbox\z@\lastbox}%
      \fi
    \else
      \clubpenalty 1
      \everypar{}%
    \fi}}
\makeatother

\usepackage[colorlinks=true, linktoc=all, citecolor=blue, urlcolor=blue, linkcolor=blue, hypertexnames=false]{hyperref}
\usepackage[capposition=top]{floatrow}
\usepackage{amsmath}
\usepackage{pxfonts}
\usepackage{amssymb,mathtools}
\usepackage{amsfonts}

\usepackage{amsthm}
\usepackage{algorithmic}
\usepackage{graphicx}
\usepackage{textcomp}
\usepackage{xcolor}
\usepackage[misc]{ifsym}
\usepackage{graphics}
\usepackage{graphicx}
\usepackage{enumitem}
\usepackage{xspace}
\usepackage{caption}
\usepackage[labelformat=parens]{subcaption}
\usepackage{amsmath}
\usepackage{amssymb}
\usepackage{bm}
\usepackage{array}
\usepackage{textcomp}
\usepackage{weiwAlgorithm}
\usepackage{url}
\usepackage{colortbl}
\usepackage{balance}
\usepackage{xparse}
\usepackage{multirow}
\usepackage{booktabs}
\usepackage{balance}
\usepackage{lipsum}

\definecolor{rehigh}{rgb}{1.0, 0.49, 0.0}
\definecolor{jiadong}{rgb}{0, 0, 0}
\NewDocumentCommand\rehigh{m}{\textcolor{rehigh}{#1}}
\NewDocumentCommand\jiadong{m}{\textcolor{jiadong}{#1}}

\newcommand{\revise}[1]{{#1}}

\newtheorem{example}{\textbf{Example}}
\newtheorem{theorem}{\textbf{Theorem}}
\newtheorem{corollary}{Corollary}
\newtheorem{definition}{\textbf{Definition}}
\newtheorem{lemma}{\textbf{Lemma}}

\DeclareMathOperator*{\argmax}{arg\,max}

\begin{document}
\title{Influence Minimization via Blocking Strategies\footnote{A preliminary study of this manuscript~\citep{0002ZW0023} was published in the Proceedings of 39th IEEE International Conference on Data Engineering (IEEE ICDE 2023), which only focuses on the problem under the node blocking strategy and the IC model.}}

\author{Jiadong Xie\thanks{The Chinese University of Hong Kong (email: jdxie@se.cuhk.edu.hk)}~~~~~~Fan Zhang\thanks{Guangzhou University (email: zhangf@gzhu.edu.cn)}~~~~~~Kai Wang\thanks{Antai College of Economics and Management, Shanghai Jiao Tong University (e-mail: w.kai@sjtu.edu.cn)}~~~~~~Jialu Liu\thanks{Antai College of Economics and Management, Shanghai Jiao Tong University (e-mail: jialuliu@sjtu.edu.cn)}\\
Xuemin Lin\thanks{Antai College of Economics and Management, Shanghai Jiao Tong University (e-mail: xuemin.lin@sjtu.edu.cn)}~~~~~~~~Wenjie Zhang\thanks{University of New South Wales (e-mail: wenjie.zhang@unsw.edu.au)}}

\date{}

\maketitle

\begin{abstract}

\noindent We study the influence minimization problem: given a graph $G$ and a seed set $S$, blocking at most $b$ nodes or $b$ edges such that the influence spread of the seed set is minimized. 
This is a pivotal yet underexplored aspect of network analytics, which can limit the spread of undesirable phenomena in networks, such as misinformation and epidemics. 
Given the inherent NP-hardness of the problem under the independent
cascade (IC) and linear threshold (LT) models, previous studies have employed greedy algorithms and Monte Carlo Simulations for its resolution. However, existing techniques become cost-prohibitive when applied to large networks due to the necessity of enumerating all the candidate blockers and computing the decrease in expected spread from blocking each of them. This significantly restricts the practicality and effectiveness of existing methods, especially when prompt decision-making is crucial. In this paper, we propose the AdvancedGreedy algorithm, which utilizes a novel graph sampling technique that incorporates the dominator tree structure. 
We find that AdvancedGreedy can achieve a $(1-1/e-\epsilon)$-approximation in the problem under the LT model.
For the problem under the IC model, since the problem is APX-hard, we further propose a novel heuristic algorithm GreedyReplace, based on identifying the relationships among candidate blockers. Experimental evaluations on real-life networks reveal that our proposed algorithms exhibit a significant enhancement in efficiency, surpassing the state-of-the-art algorithm by three orders of magnitude, while achieving high effectiveness. 




\end{abstract}


\clearpage

\input{MS_Version/intros_MS}
\input{MS_Version/relates_MS}
\input{preli.tex}

\input{analy.tex}

\input{base.tex}
\input{estim.tex}

\input{imin.tex}
\input{MS_Version/exp_MS}

\section{Conclusion} \label{ses:conclusion}
\jiadong{This study focuses on the influence minimization problem, specifically the selection of a blocker set consisting of $b$ nodes or $b$ edges to minimize the influence spread from a given seed set. This problem is critical yet relatively understudied, with the potential to tackle urgent societal challenges such as combating misinformation spreads and managing epidemic outbreaks. A major obstacle in influence minimization has been the absence of efficient and scalable techniques suitable for handling large networks.
Existing approaches often struggle with the computational burden posed by real-life large networks, which hinders their practicality and effectiveness. Through meticulous analysis, we have successfully proven the NP-hardness of the influence minimization problem. Additionally, this study introduces the AdvancedGreedy algorithm as a novel approach to enhance the efficiency of existing algorithms while maintaining their effectiveness, surpassing them by a remarkable three orders of magnitude. The efficiency of our algorithm is enhanced by simultaneously computing the reduction in expected influence spread using sampled graphs and their dominator trees. Furthermore, we introduce the GreedyReplace algorithm as an enhancement to the greedy method. This algorithm incorporates a new heuristic, resulting in improved effectiveness. Our proposed AdvancedGreedy and GreedyReplace algorithms have been extensively tested on $8$ real-life datasets, and the results confirm their superior performance compared to other existing algorithms in the literature. In summary, our proposed algorithm's unprecedented speedup marks a turning point in the feasibility of influence minimization, which empowers timely decision-making in scenarios where rapid intervention is paramount.}

\jiadong{
Despite the contributions of our paper, it is important to discuss the limitations and future directions of our research. First, this study assumes a static network, but the algorithms proposed here can serve as inspiration for developing efficient and effective algorithms for dynamic networks. Additionally, future research can explore efficient parallel and distributed solutions based on the proposed algorithms. Furthermore, it would be worthwhile to investigate more efficient heuristics that provide satisfactory trade-offs in terms of result quality.}


\bibliographystyle{informs2014}
\bibliography{ref}

\newpage
\input{MS_Version/Appendix_MS}

\end{document}

%% file: MS_Version/intros_MS.tex
\section{Introduction}\label{sec:intros}

\jiadong{
In today's hyper-connected world, the concept of influence maximization, which aims to target a select group of individuals in order to maximize the spread of influence \citep{value-customer, first-max}, has long been a focal point of research due to its commercial potential in various industries such as marketing, finance, education, etc \citep[e.g.,][]{maximization1, banerjee2013diffusion, gunnecc2020least, wu2020diffnet++,raghavan2022influence,eckles2019seeding,zhou2023detecting,TengXZWZ23}. However, a lesser-explored yet equally critical counterpart to this problem is the \emph{influence minimization} problem. This problem involves identifying a specific set of nodes (or edges) in a network in order to minimize the spread of influence, particularly when some forms of influence are detrimental. Let us provide the following two examples to illustrate the importance of minimizing the spread of influence.}



\noindent
\jiadong{
\textbf{Reducing Misinformation Propagation.} In the era of rapid information dissemination and the prevalence of social media, the proliferation of misinformation has emerged as a major societal concern and is regarded as one of the most significant global economic risks \citep{WEF2022}. The dissemination of false information or unverified rumors can lead to various adverse effects, such as economic damage, substantial disruption, and widespread panic. For example, a false tweet claiming that Barack Obama was injured in an explosion at the White House led to a significant decline of \$136.5 billion in the stock market~\citep{fakenews}. 
Therefore, it is crucial to identify influential nodes or edges within the social network to restrict the dissemination of such misinformation and mitigate the potential negative consequences.}

\noindent \jiadong{\textbf{Enhancing Epidemic Control.} The outbreak of infectious diseases poses formidable challenges to public health systems worldwide, demanding swift and strategic intervention to limit transmission. 
For example, COVID-19 is being attributed as the final factor leading to the breakdown of the United Kingdom's healthcare system \citep{mclellan2022nhs}. In such scenarios, timely responses become paramount to mitigate the potential cascading effects on morbidity, mortality, and socio-economic stability. The adverse effects of infectious disease outbreaks underscore the importance of harnessing innovative solutions that can efficiently identify critical nodes or edges within transmission networks and guide swift containment measures.}

\jiadong{
The above examples highlight the urgent need for effective solutions to the influence minimization problem. The problem can be formulated as follows \citep{min-greedy1}: a social network (or a transmission network) is modeled as a graph where nodes represent individuals and edges represent connections or relationships between them. The propagation of influence (e.g., fake news or diseases) in a network follows a diffusion model, such as the \emph{independent cascade} (IC) model or \emph{linear threshold} (LT) model~\citep{first-max}. Given a graph $G$, a seed set $S$, and a budget $b$, the objective of the influence minimization problem is to identify a set $B^*$ consisting of at most $b$ nodes (or $b$ edges) that minimizes the expected number of nodes activated by the seeds $S$ (referred to as the expected influence spread) under a specific diffusion model.}

\jiadong{
Existing research has explored various methods to tackle the influence minimization problem \citep[e.g.,][]{min-edge1,min-greedy1, min-edge2,min-greedy-tree}. However, a common limitation observed in these approaches is their high time complexity. For instance, the processing time required for the Wiki-Vote network (7K nodes and 103K edges) exceeds three days when using these methods. This significantly hinders their applicability to large networks and real-time scenarios where prompt decision-making is essential for effective intervention \citep{DBLP:journals/jnca/ZareieS21}. From a pragmatic perspective, it is impractical for an algorithm to require a longer duration to discover a solution that minimizes influence spread compared to the actual time required for influence to propagate. }

\jiadong{
To address the limitation, we present novel and efficient algorithms that tackle the influence minimization problem. Our algorithms address the scalability issues encountered by existing solutions, providing a significant leap forward in terms of computational speed and adaptability to large-scale real-world networks. Our proposed algorithms showcase an unprecedented speedup, outperforming existing methods by an astonishing three orders of magnitude. In addition, we conduct an exhaustive series of real-data experiments to evaluate the efficacy and applicability of our proposed algorithms in a variety of scenarios. In the following section, we discuss the various technical challenges encountered and the corresponding contributions in detail.}

\vspace{1mm}
\noindent \textbf{Main Technical Challenges.}
\jiadong{
There are several technical challenges in influence minimization problems. First, the influence minimization problem is NP-hard under both node blocking and edge blocking strategies with the IC/LT influence propagation model (Theorems~6 and 7). Then, the state-of-the-art algorithms employ a greedy framework to select the blockers \citep{min-greedy1,min-greedy2}, which has proven to be superior to alternative heuristics proposed in previous studies \citep[e.g,][]{Nature-error,Viruses,min-greedy-tree}. However, in contrast to the influence maximization problem, we demonstrate that the expected spread function in the influence minimization problem does not exhibit supermodularity (Theorem~8). As a result, existing greedy solutions may not possess an approximation guarantee.}

\jiadong{Second, similar to the influence maximization problem, an essential component of the greedy algorithm involves the computation of the influence spread from the seeds, a problem known to be \#P-hard~\citep{maximization1}. Rather than relying on an exact algorithm, state-of-the-art solutions for estimating influence spread in diffusion models utilize Monte-Carlo Simulations \citep{influencepath,OhsakaAYK14,MS-PCG}.
In contrast to the influence maximization problem, the influence minimization problem incorporates an additional step, which involves the computation of the reduction in expected influence spread that would occur by blocking a particular node or edge (detailed in Section~\ref{sec:ec-exist}). In order to accomplish this, the current algorithms must iterate through all potential nodes or edges. Given the presence of an overwhelming number of potential candidate nodes or edges that could be blocked, the implementation of these methods becomes economically unfeasible when applied to larger graphs.}

\vspace{1mm}
\noindent \textbf{Our Contributions.}
\jiadong{
In this paper, we prove that the influence minimization problem is NP-hard under different blocking strategies (node blocking or edge blocking) with various influence propagation models (IC or LT). For the problem under the LT model, we find that the expected spread function is supermodular, which enables us to leverage the greedy algorithm to achieve a $(1-1/e-\epsilon)$-approximation.}
\jiadong{We then tackle the issue of scalability by introducing a novel algorithm, namely the AdvancedGreedy algorithm. This algorithm exhibits a remarkable increase in speed by several orders of magnitude compared to all preexisting greedy algorithms while maintaining their efficacy. The efficiency of our algorithm is enhanced through the simultaneous computation of the expected influence spread reduction for each candidate node {and edges}. This is achieved by employing an almost linear scan of each sampled graph.
{The central concept underlying our algorithm is the utilization of the dominator tree for the computation of the expected reduction in expected influence spread. This novel approach has not yet been employed in the context of influence-related problems.} We prove that the decrease of the expected influence spread from a blocked node is decided by the subtrees rooted in the dominator trees that are generated from the sampled graphs (Theorem~\ref{theo:dominator-subtree}), and we establish that the edge version problem can be effectively solved by transforming it into the node version problem (Theorems~\ref{theo::edge_expected} and \ref{theo:edge-dominator-subtree}).
Hence, rather than relying on Monte Carlo Simulations, we can effectively calculate the expected reduction in influence spread by employing sampled graphs and their dominator trees. We also prove the estimation result is theoretically guaranteed given a certain number of samples (Theorem~\ref{theorem:approx}). Furthermore, we prove the AdvancedGreedy algorithm can achieve a $(1-1/e-\epsilon)$-approximation for both node and edge blocking problems under the LT model.}

\jiadong{
In addition, for the problem under the IC model, since the problem is hard to approximate, we further propose a novel heuristic. When employing the node-blocking strategy, it is evident that if the budget is unlimited, all out-neighbors of the seeds will be blocked. However, the current greedy algorithm may select nodes that are not the out-neighbors as blockers, potentially overlooking crucial candidates. Considering this, we propose a new heuristic, named GreedyReplace algorithm, based on identifying the relationships among candidate blockers. Specifically, we begin by restricting the pool of candidate blockers to the out-neighbors of the blocking nodes. Subsequently, we adopt a greedy strategy to replace these blockers with alternative nodes, should the expected influence spread diminish. This algorithm demonstrates superior performance. }

\jiadong{
We conduct extensive experiments on $8$ real-life publicly available datasets with varying scales, features, and diffusion models. Our proposed algorithms are compared with the state-of-the-art algorithm, namely the greedy algorithm with Monte-Carlo Simulations \citep{min-greedy2,min-greedy1,min-greedy-tree}, as well as other heuristics. Our experiments show that (a) our AdvancedGreedy algorithm exhibits significantly enhanced speed when compared to existing algorithms, surpassing them by a factor of more than three orders of magnitude, (b) our GreedyReplace algorithm demonstrates superior effectiveness, as evidenced by its ability to achieve smaller influence spreads while maintaining comparable efficiency to our AdvancedGreedy algorithm. Our algorithms demonstrate remarkable scalability, enabling it to easily apply to networks comprising millions of nodes and edges.}


\jiadong{Overall, our research makes significant strides in the realm of influence minimization problems, providing innovative solutions that overcome existing limitations. Our proposed algorithms outperform existing methods by several orders of magnitude in terms of speed. By achieving such remarkable computational efficiency, our approach opens new avenues for managing large-scale networks and real-time scenarios, enabling faster and more proactive responses to potential crises.}

\jiadong{
The remainder of this paper is structured as follows: In Section \ref{sec:relate}, we present a comprehensive review of the related literature. Section \ref{sec:pre} offers a formal definition of the influence minimization problem, introduces the most advanced existing algorithm, and establishes the NP-hard nature of the influence minimization problem. Section \ref{sec::decrease_spread} provides a detailed description of how we calculate the reduction of expected influence spread. Section \ref{sec::app-IMIN} introduces our proposed algorithms and their implementation. Section \ref{sec:exp} outlines the experiment and presents the results obtained from applying our approach to real-life graphs. 
Finally, Section \ref{ses:conclusion} concludes this work, summarizing the main contributions and discussing future research directions.}

A preliminary study of this paper~\citep{0002ZW0023} was presented in the Proceedings of 39th IEEE International Conference on Data Engineering (IEEE ICDE 2023), which only focuses on the problem under the node blocking strategy and the IC model.

%% file: MS_Version/relates_MS.tex
\section{\rehigh{Related Works}}
\label{sec:relate}

\noindent \rehigh{\textbf{The Influence Maximization Problem.}
There are many studies of influence maximization
\citep{DBLP:journals/kais/BanerjeeJP20,DBLP:journals/computing/AghaeeGBBF21}.
Domingos and Richardson are the first to study the influence between individuals for marketing in social networks \citep{value-customer}. 
Later, this problem was formulated as a discrete optimization problem, referred to as the influence maximization problem \citep{first-max}.
They introduce the IC and LT diffusion models and propose a greedy algorithm that achieves a $(1-1/e)$-approximation ratio due to the submodularity of the function under these models.
Subsequently, some researchers have proposed advanced approaches, such as reverse sampling \citep{Borgs-max}.
Other researchers have explored strategies to maximize influence in practical scenarios that may involve domain-specific challenges, e.g., costly network information \citep{eckles2019seeding}, user countering \citep{XieCCZLT24}.}



\rehigh{In contrast to prior research, the objective of this paper is to contribute to the influence minimization problem field, which presents unique challenges that cannot be addressed using existing solutions for influence maximization problem. While the influence minimization problem may appear to be the inverse of the influence maximization problem, it actually entails additional complexities. Specifically, it requires the estimation of the expected reduction in influence spread resulting from the blocking of nodes or edges.}

\noindent \rehigh{\textbf{The Influence Minimization Problem.}
Compared with the influence maximization problem, there are fewer studies on minimizing the spread of influence \citep{DBLP:journals/jnca/ZareieS21}. In general, there are two main strategies that can be employed to solve the influence minimization problem. The first strategy involves blocking (or removing) specific nodes or edges to minimize the overall impact of influence \citep[e.g.,][]{min-edge1,min-greedy1,YaoSZWG15,min-greedy-tree}. The second strategy, known as the counterbalance strategy, involves the dissemination of positive information throughout the network \citep[e.g.,][]{min-greedy2,LeeSMH19,seed-positive,ManouchehriHD21}. By increasing people's awareness and providing them with positive content, the negative influence is reduced. This methodology is predominantly employed to address the dissemination of misinformation. Our papers is the first category.}

\rehigh{Motivated by the feasibility of structure change for influence-related studies \citep{struct-sn,experience-min1,user-attribute2}, most works have focused on utilizing the blocking strategy (i.e., blocking nodes or edges) to minimize the influence spread. Intuitively, simple heuristics that rely on network structure can be employed to identify influential nodes to restrict the spread of influence. For instance, one commonly used heuristic is the degree-based heuristic \citep{Nature-error,Viruses,shah2011rumors,shah2016finding}.
However, subsequent studies have shown that these simple degree-based heuristic methods are comparatively less effective than greedy heuristics \citep{min-greedy1,min-greedy-tree}. Hence, different greedy algorithms have been proposed to address the influence minimization problem by blocking nodes or edges under various diffusion models \citep{min-edge1,min-greedy1,min-edge2,min-greedy-tree}.}



\rehigh{However, a prevalent constraint observed in these methodologies is their limited ability to scale effectively when applied to large networks \citep{DBLP:journals/jnca/ZareieS21}. Even the most advanced greedy approach requires the enumeration of all candidate nodes to calculate the reduction in expected influence spread. Consequently, the computational speed of their solutions often falls short in dealing with real-time scenarios, where quick decisions are crucial for successful intervention. In this paper, we develop a computationally efficient approach that maintains its effectiveness when compared to existing algorithms for minimizing influence spread. Specifically, we consider two fundamental blocking strategies: node blocking and edge blocking. Besides, we also consider two distinct diffusion models, namely the LT model and the IC model.}

\noindent \rehigh{\textbf{The Computation of the Expected Spread.}
An essential aspect of the influence minimization problem involves the estimation of the expected spread, which has been proved to be \#P-hard under IC model \citep{maximization1}. 
Researchers first propose to use the Monte-Carlo Simulations (MCS) to compute the expected influence spread, which repeats simulations until a tight estimation is obtained \citep{first-max}. In a subsequent study \citep{Borgs-max}, researchers propose Reverse Influence Sampling (RIS). Additionally, researchers propose the first algorithm to compute the exact influence spread under the IC model \citep{MaeharaSI17}. However, its applicability is limited to small graphs containing a few hundred edges.
}

\rehigh{However, as discussed later in Section~\ref{sec:baseline}, we find that {MCS and RIS} are not applicable to our problem. This is because influence minimization problems require repetitive computation of the expected decrease in influence spread for each potential blocked node/edge. We contribute to this stream of literature by introducing the dominator tree when calculating the reduction of expected influence spread, which has not been previously applied to this type of problem.
The dominator tree has predominantly been applied in various fields, e.g., program analysis~\citep{reif1978symbolic,chen2023synthdb} and software testing~\citep{Agrawal99}, which is based on theoretical network flow problems. We, on the other hand, leverage its tree structure properties for our problem, enabling simultaneous estimation of spread reduction across all candidate nodes/edges. This approach minimizes redundant computations in contrast to MCS or RIS methods.
}


%% file: preli.tex
\section{\rehigh{Preliminaries}}
\label{sec:pre}

\rehigh{
We consider a directed graph $G=(V, E)$, where $V$ is the set of $n$ nodes (individuals), and $E$ is the set of $m$ directed edges (influence relations between node pairs). We use $V(G)$ (resp. $E(G)$) to represent the set of nodes (resp. edges) in $G$, and $p_{u,v}$ denotes the probability that node $u$ activates node $v$.
Let $G[V]$ denote the induced subgraph by the node set $V$. $N^{in}_v$ (resp. $N^{out}_v$) is the in-neighbor (resp. out-neighbor) set of $v$ in $G$.
The in-degree (resp. out-degree) of $v$ in graph $G$, is denoted by $d^{in}_v=|N^{in}_v|$ (resp. $d^{out}_v=|N^{out}_v|$).
We use $Pr[x]$ to denote the probability if $x$ is true and $\mathbb{E}[x]$ to denote the expectation of variable $x$.
Table~A1 in the online appendix summarizes the notations. }

\subsection{Diffusion Model}

\label{sec:diffmodel}
\jiadong{Following the existing studies on influence minimization~\citep{min-greedy2,min-greedy1,max-rr}, we focus on the two widely-studied diffusion models, namely IC model and LT model~\citep{first-max}, which capture the ``word-of-mouth'' effects, i.e., each node’s tendency to become active increases monotonically as more of its neighbors become active.
They both assume each directed edge $(u,v)$ in the graph $G$ has a propagation probability
$p_{u,v}\in [0,1]$, i.e., the probability that the node $u$ activates the node $v$ after $u$ is activated. In both two diffusion models, each node has two states: inactive or active. We say a node is activated if it becomes active by the influence spread (e.g., fake news or diseases), and an active node will not be inactivated during the diffusion process.}



\noindent \jiadong{\textbf{LT model.}
LT model is proposed to capture the word-of-mouth effects based on the use of node-specific thresholds \citep{granovetter1978threshold,schelling2006micromotives}.
In LT model, each node $v$ independently selects a threshold $\theta_v$ from a uniform distribution over the range $[0, 1]$. The model considers an influence propagation process as follows: (i) at timestamp $0$, the seed nodes are activated, and the other nodes are inactive; (ii) at any timestamp $t>0$, an inactive node $v$ is activated if and only if the total weight of its active in-neighbors exceeds its threshold $\theta_v$ (i.e., $\sum_{u\in (S'\cup N^{in}_v)}p_{u,v}\ge \theta_v$, where $S'\subseteq V$ is the set of active nodes at timestamp $t-1$); and (iii) we repeat the above 
steps until no node can be activated at the latest timestamp.}


\noindent \rehigh{\textbf{IC model.}
Based on work in interacting particle systems from probability theory \citep{liggett1985interacting}, the IC model is proposed to consider dynamic cascade models for diffusion processes.
The model considers an influence propagation process as follows: (i) at timestamp $0$, the seed nodes are activated, i.e., the seeds are now active while the other nodes are inactive; (ii) if a node $u$ is activated at timestamp $i$, then for each of its inactive out-neighbor $v$ (i.e., for each inactive $v\in N^{out}_u$), $u$ has $p_{u,v}$ probability to independently activate $v$ at timestamp $i+1$; and (iii) we repeat the above steps until no node can be activated at the latest timestamp.}


\subsection{Problem Definition and Hardness}
\label{sec:define}
\rehigh{To formally introduce the influence minimization problem \citep{min-greedy1,min-greedy-tree}, we first define the node activation probability, which is initialized by $1$ for any seed node by default.}

\vspace{-5mm}
\rehigh{\begin{definition}[activation probability]
Given a directed graph $G$, a node $x$ and a seed set $S$, the activation probability of $x$ in $G$, denoted by $\mathcal{P}^G(x,S)$, is the probability of the node $x$ becoming active. 
\end{definition}}

\rehigh{
In order to minimize influence spread, we can block some non-seed nodes or edges such that they will not be activated in the propagation process. In this paper, a \textit{blocked node} is also called a \textbf{blocker}, that is, the influence probability of every edge pointing to a blocker is set to $0$.
\revise{The activation probability of a blocker is $0$ because the propagation probability is $0$ for any of its incoming edges.} For blocking edges, the influence probability of a \textit{blocked edge} is set to $0$.}

\rehigh{
Then, we define the expected spread to measure the influence of the seed set in the whole graph. We use the terms ``expected influence spread'' and ``expected spread'' interchangeably for the sake of brevity.}


\vspace{-5mm}
\rehigh{
\begin{definition}[expected spread]
Given a directed graph $G$ and a seed set $S$, the expected spread, denoted by $\mathbb{E}(S,G)$, is the expected number of active nodes, i.e., $\mathbb{E}(S,G)=\sum_{u\in V(G)}\mathcal{P}^G(u,S)$.
\end{definition}}


\rehigh{The expected spread with a blocker set $B$ is represented by $\mathbb{E}(S,G[V\setminus B])$. Recall that there exist two blocking strategies for addressing the influence minimization problem, namely node blocking (referred to as IMIN problem) and edge blocking (referred to as IMIN-EB problem) strategies.}


\vspace{-5mm}
\rehigh{
\begin{definition}[IMIN problem]
Given a directed graph $G=(V,E)$, the influence probability $p_{u,v}$ on each edge $(u,v)$, a seed set $S$ and a budget $b$, the influence minimization problem is to find a blocker set $B^*$ with at most $b$ nodes such that the influence (i.e., expected spread) is minimized, i.e., 
$B^*={\arg \min}_{B\subseteq (V\setminus S),|B|\le b} \mathbb{E}(S,G[V\setminus B]).$
\end{definition}}




\vspace{-5mm}
\jiadong{\begin{definition}[IMIN-EB problem]
Given a directed graph $G=(V,E)$, the influence probability $p_{u,v}$ on each edge $(u,v)$, a seed set $S$ and a budget $b$, the Influence Minimization via Edge Blocking (IMIN-EB) problem is to find an edge set $B^*$ with at most $b$ edges such that the influence (i.e., expected spread) is minimized, i.e., 
$B^*={\arg \min}_{B\subseteq E,|B|\le b} \mathbb{E}(S,G(V, E\setminus B)).$
\end{definition}}

\begin{figure}[!htbp]
    \centering
        \centering    
        \includegraphics[width=.3\columnwidth]{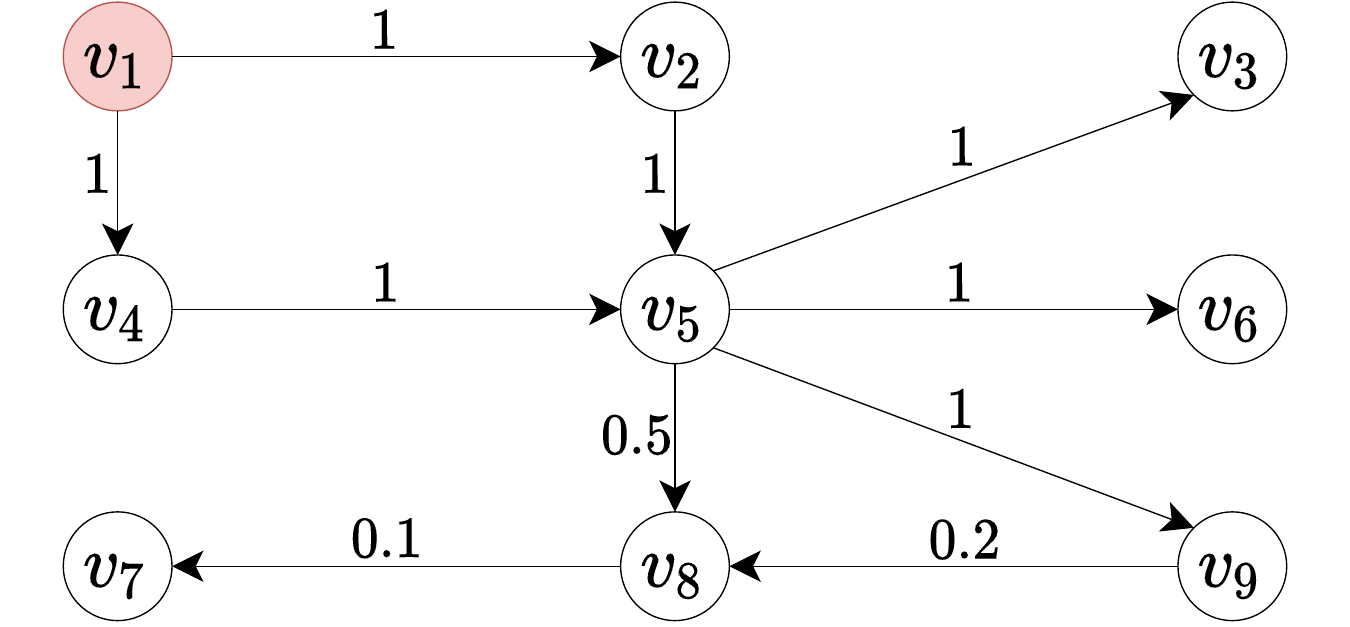}
        
    \caption{A toy graph $G$, where $v_1$ is the seed node and the value on each edge indicates its propagation probability}
    \vspace{-3mm}
\label{fig:sample}
\end{figure}
\begin{example}
\label{example:compute} \normalfont
Figure~\ref{fig:sample} shows a graph $G = (V, E)$ where $S=\{v_1\}$ is the seed set, and the value on each edge is its propagation probability, e.g., $p_{v_5,v_8}=0.5$ indicates $v_8$ can be activated by $v_5$ with $0.5$ probability if $v_5$ becomes active. At timestamp $0$, seed $v_1$ node is activated while other nodes are inactive. At timestamps $1$ to $3$, the seed $v_1$ will certainly activate $v_2,v_3,v_4,v_5,v_6$ and $v_9$, as the corresponding activation probability is $1$. 
Because $v_8$ may be activated by either $v_5$ or $v_9$, we have $\mathcal{P}^G(v_8,\{v_1\})=1-(1-p_{v_5,v_8}\cdot\mathcal{P}^G(v_5,\{v_1\}))(1-p_{v_9,v_8}\cdot\mathcal{P}^G(v_9,\{v_1\}))=1-0.5\times 0.8 = 0.6$. 
If $v_8$ is activated, it has $0.1$ probability to activate $v_7$.
Thus, we have $\mathcal{P}^G(v_7,\{v_1\})=p_{v_8,v_7}\cdot\mathcal{P}^G(v_8,\{v_1\})=0.6\times 0.1=0.06$. 
The expected spread is the activation probability sum of all the nodes, i.e., $\mathbb{E}(\{v_1\},G)=7.66$. 
In IMIN problem, if we block $v_5$, the new expected spread $\mathbb{E}(\{v_1\},G[V\setminus \{v_5\}])=3$. Similarly, we have $\mathbb{E}(\{v_1\},G[V\setminus \{v_2\}])=\mathbb{E}(\{v_1\},G[V\setminus \{v_4\}])=6.66$, and blocking any other node also achieves a smaller expected spread than blocking $v_5$. Thus, if $b = 1$, the result of the IMIN problem is $\{v_5\}$.
\jiadong{For IMIN-EB problem, if we block $(v_5,v_9)$, the expected spread will become $\mathbb{E}(\{v_1\},G(V,E\setminus \{(v_5,v_9)\}))=6.55$. Similarly, we have $\mathbb{E}(\{v_1\},G(V,E\setminus \{(v_5,v_3)\}))=6.66$, and blocking any other edge also achieves a smaller expected spread than blocking $(v_5,v_9)$. Thus, if the budget $b = 1$, the result of the IMIN-EB problem is $\{(v_5,v_9)\}$.}
\end{example}

%% file: analy.tex
{\noindent\textbf{Problem Hardness.}
We prove that both the IMIN and IMIN-EB problems are NP-Hard under the IC/LT model, as demonstrated in Theorems~6 and~7. The detailed proofs can be found in online appendix 1.}

%% file: base.tex
\subsection{A Baseline Algorithm}
\label{sec:baseline}

\jiadong{In this subsection, we review and discuss the baseline greedy algorithm, which is the most advanced existing solution to IMIN and IMIN-EB problem~\citep{min-greedy2,min-greedy1,PhamPHPT19,min-greedy-tree,acycle-min-greedy}. This algorithm serves as a benchmark for comparison with our proposed algorithms.}

\jiadong{The baseline greedy algorithm is as follows (psuedo-code in online appendix 2): we start with an empty blocker set (resp. edge set) $B=\varnothing$, and then iteratively add node $u$ (resp. edge $(u,v)$) into set $B$ that leads to the largest decrease of expected spread, i.e., $u=\arg \max _{u\in V\setminus (S\cup B)}(\mathbb{E}(S,G[V\setminus B])-\mathbb{E}(S,G[V\setminus (B\cup \{u\})]))$ (resp. $(u,v)=\arg \max _{(u,v)\in E\setminus B}(\mathbb{E}(S,G(V,E\setminus B))-\mathbb{E}(S,G(V, E\setminus (B\cup \{(u,v)\}))))$), until $|B|=b$.}



\jiadong{As the greedy algorithms proposed by previous works use Monte-Carlo Simulations to compute the expected spread, each computation of spread decrease needs $O(r\cdot m)$ time, where $r$ is the number of rounds in Monte-Carlo Simulations. Thus, the time complexity of the baseline greedy algorithm is $O(b\cdot n\cdot r\cdot m)$ for the IMIN problem (resp. $O(b\cdot m\cdot r\cdot m)$ for the IMIN-EB problem).}
\rehigh{As indicated by the complexity, the baseline greedy algorithm cannot efficiently handle the cases with large $b$. 
The greedy heuristic is usually effective on small $b$ values. However, it still incurs a significant time cost due to the need to enumerate the entire node set as candidate blockers and compute the expected spread for each candidate.}

%% file: estim.tex
\section{Computation of Decrease in Expected Spread}
\label{sec::decrease_spread}

\jiadong{In this section, we initially review the existing approaches for the influence minimization problem and recognize that estimating the expected influence spread plays a pivotal role in enhancing algorithmic efficiency (Section~\ref{sec:baseline}). However, the current techniques for computing the expected spread are not feasible for efficiently estimating the reduction in expected spread caused by each candidate blocker (Section~\ref{sec:baseline}). Consequently, we introduce an algorithm designed to efficiently assess the reduction in expected spread resulting from the blocking of a single node within the graph.
Specifically, we put forth a novel framework (Section~\ref{sec:new-es}) that leverages sampled graphs (Section~\ref{sec::sample-graph}) and their dominator trees (Section~\ref{sec::domi-tree}) to effectively compute the reduction in expected spread for each potential blocker. This computation can be performed with a single scan of the dominator trees.
In addition, our approach can be utilized to estimate the reduction in the expected spread that arises from edge blocking by adding virtual nodes in the graph (Section~\ref{sec:new-es}).}

{\color{rehigh}
\label{sec:onesource}
\noindent
\textbf{From Multiple Seeds to One Seed.}
For simplicity in presentation, we introduce the techniques for the case of one seed node.
A unified seed node $s'$ is created to replace all the seeds in the graph. For each node $u$, if there are $h$ different seeds pointing to $u$ and the probability on each edge is $p_i (1\le i\le h)$, we remove all the edges from the seeds to $u$ and add an edge from $s'$ to $u$ with probability $(1-\prod_{i=1}^h (1-p_i))$ under the IC model (resp. $\sum _{i=1}^h p_i$ under the LT model).

{\subsection{A Baseline Greedy Approach and Existing Expected Spread Estimation Techniques}
\label{sec:baseline}

\jiadong{We first review and discuss the baseline greedy algorithm, which is the most advanced existing solution to IMIN and IMIN-EB problem~\citep{min-greedy2,min-greedy1,PhamPHPT19,min-greedy-tree,acycle-min-greedy}. This algorithm serves as a benchmark for comparison with our proposed algorithms.}
\jiadong{The baseline greedy algorithm is as follows (psuedo-code in online appendix 2): we start with an empty blocker set (resp. edge set) $B=\varnothing$, and then iteratively add node $u$ (resp. edge $(u,v)$) into set $B$ that leads to the largest decrease of expected spread, i.e., $u=\arg \max _{u\in V\setminus (S\cup B)}(\mathbb{E}(S,G[V\setminus B])-\mathbb{E}(S,G[V\setminus (B\cup \{u\})]))$ (resp. $(u,v)=\arg \max _{(u,v)\in E\setminus B}(\mathbb{E}(S,G(V,E\setminus B))-\mathbb{E}(S,G(V, E\setminus (B\cup \{(u,v)\}))))$), until $|B|=b$.}



\jiadong{As the greedy algorithms proposed by previous works use Monte-Carlo Simulations to compute the expected spread, each computation of spread decrease needs $O(r\cdot m)$ time, where $r$ is the number of rounds in Monte-Carlo Simulations. Thus, the time complexity of the baseline greedy algorithm is $O(b\cdot n\cdot r\cdot m)$ for the IMIN problem (resp. $O(b\cdot m\cdot r\cdot m)$ for the IMIN-EB problem).}
\rehigh{As indicated by the complexity, the baseline greedy algorithm cannot efficiently handle the cases with large $b$. 
The greedy heuristic is usually effective on small $b$ values. However, it still incurs a significant time cost due to the need to enumerate the entire node set as candidate blockers and compute the expected spread for each candidate.}

Since computing the expected influence spread of a seed set in both IC and LT model is \#P-hard \citep{maximization1}, and the exact solution can only be used in small graphs (e.g., with a few hundred edges) \citep{MaeharaSI17}, previous research focuses on estimating the expected influence spread, {rather than exact computation}.
There are two existing methods for estimating the expected spread.}
\label{sec:ec-exist}


\noindent \textbf{Monte-Carlo Simulations (MCS).}
Kempe et al. \citep{first-max} apply MCS to estimate the influence spread, which is often used in some influence-related problems \citep[e.g.,][]{influencepath,OhsakaAYK14,MS-PCG}.
In each round of MCS: (a) for the IC model, it removes every edge $(u,v)$ with $(1-p_{u,v})$ probability; (b) for the LT model, for each node $v$, assume it has $k$ incoming edges $(u_1,v),\cdots,(u_k,v)$, $v$ has $p_{u_i,v} (i\in [1,k])$ probability to have a unit incoming edge $(u_i,v)$ and $1-\sum_{i=1}^k p_{u_i,v}$ probability to have no incoming edge.
Let $G'$ be the resulting graph, and the set $R(s)$ contains the nodes in $G'$ that are reachable from $s$ (i.e., there exists at least one path from $s$ to each node in $R(s)$). For the original graph $G$ and seed $s$, the expected size of set $R(s)$ equals the expected spread $\mathbb{E}(\{s\},G)$ \citep{first-max}. 
Assuming we take $r$ rounds of MCS to estimate the expected spread, MCS needs $O(r\cdot m)$ times to calculate the expected spread. 
Recall that the IMIN (resp. IMIN-EB) problem is to find the optimal node (resp. edge) set with a given seed set.
The spread computation by MCS for the two problems is costly. This is because the dynamic of influence spread caused by different nodes (resp. edges) is not fully utilized in the sampling. Consequently, it becomes necessary to repeatedly perform MCS for each candidate set.

\noindent \textbf{Reverse Influence Sampling (RIS).}
Borgs et al.~\citep{Borgs-max} propose RIS to approximately estimate the influence spread, which is often used in the solutions for influence maximization (IMAX) problem \citep[e.g.,][]{SunHYC18,Guo0WC20}. For each round, RIS generates an instance of $g$ randomly sampled from graph $G$ in the same way as MCS. Then a node $x$ is randomly sampled in $g$. It performs reverse breadth-first search (BFS) to compute the reverse reachable (RR) set of the node $x$, i.e., the nodes that can be reached by node $x$ in the reverse graph of $g$. They prove that if the RR set of node $v$ has $\rho$ probability of containing the node $u$ when $u$ is the seed node, we have $\rho$ probability of activating $v$.
In IMAX problem, RIS generates RR sets by sampling the nodes in the sampled graphs and then applying the greedy heuristic. 
As the expected influence spread is submodular of seed set~\citep{expectedspread}, an approximation ratio can be guaranteed by RIS in IMAX problem.
{However, for the influence minimization problem, reversing the graph is not helpful. This is because the blockers (resp. edges which are blocked) appear to act as ``intermediaries" between the seeds and other nodes. Hence, the computation cannot be unified into a single process in the reversing.}}

\subsection{Expected Spread Estimation via Sampled Graphs} \label{sec::sample-graph}

\jiadong{We first introduce the randomly sampled graph under the IC and LT models.}

\noindent \jiadong{\textbf{IC model.}
Let $\mathcal{G}$ be the distribution of the graphs with each induced by the randomness in edge removals from $G$, i.e., removing each edge $e=(u,v)$ in $G$ with $1-p_{u,v}$ probability. A random sampled graph $g$ derived from $G$ is an instance randomly sampled from $\mathcal{G}$.}

\begin{table}[t]
    \caption{Notations for Random Sampled Graph}
    \small
    \label{tab:nota-sampled}
       \centering
       \resizebox{0.9\linewidth}{!}{
       \begin{tabular}{|p{0.1\columnwidth}|p{0.85\columnwidth}|}\hline
        \revise{Notation} & \revise{Definition}\\ \hline\hline
       \revise{$\sigma(s,G)$} &
       \revise{the number of nodes reachable from $s$ in $G$}\\ \hline
       \revise{$\sigma^{\rightarrow u}(s,G)$} &
\revise{the number of nodes reachable from $s$ in $G$, where all the paths from $s$ to these nodes pass through $u$}
       \\ \hline
       \revise{$\xi^{\rightarrow u}(s,G)$} & \revise{the average number of $\sigma^{\rightarrow u}(s,G)$ in the sampled graphs which are derived from $G$}\\ \hline
       \end{tabular}}
\end{table}

\noindent \jiadong{\textbf{LT model.}
Let $\mathcal{G}$ be the distribution of the graphs with each induced by the randomness in incoming edge selections from $G$, i.e., for each node $v$, assume it has $k$ incoming edges $(u_1,v),\cdots,(u_k,v)$, $v$ has $p_{u_i,v} (i\in [1,k])$ probability to have a unit incoming edge $(u_i,v)$ and $1-\sum_{i=1}^k p_{u_i,v}$ probability to have no incoming edge. A random sampled graph $g$ derived from $G$ is an instance randomly sampled from $\mathcal{G}$.}

{We summarize the notations related to the randomly sampled graph in Table~\ref{tab:nota-sampled}.}


{\color{rehigh}The following lemma is a useful interpretation of expected spread~\citep{MaeharaSI17}.

\vspace{-2mm}
\begin{lemma}
\label{lemma:sample}
Suppose that the graph $g$ is a randomly sampled graph derived from $G$. Let $s$ be a seed node, we have $\mathbb{E}[\sigma(s,g)]=\mathbb{E}(\{s\},G)$.
\end{lemma}

\revise{By Lemma~\ref{lemma:sample}, we have the following corollary for computing the expected spread.}

\vspace{-2mm}
\begin{corollary}
\label{corollary::es-onenode}
\revise{Given two fixed nodes $s$ and $u$ with $s\neq u$, and a random sampled graph $g$ derived from $G$, we have $\mathbb{E}[\sigma(s,g)-\sigma^{\rightarrow u}(s,g)]=\mathbb{E}(\{s\},G[V\setminus \{u\}])$.}
\end{corollary}


\revise{Based on Lemma~\ref{lemma:sample} and Corollary~\ref{corollary::es-onenode}, we can compute the decrease of expected spread.}

\vspace{-2mm}
\begin{theorem}
Let $s$ be a fixed node, $u$ be a blocked node, and $g$ be a randomly sampled graph derived from $G$, respectively. For any node $u\in V(G)\setminus \{s\}$, we have the decrease of expected spread by blocking $u$ is equal to $\mathbb{E}[\sigma^{\rightarrow u}(s,g)]$, where $\mathbb{E}[\sigma^{\rightarrow u}(s,g)]=\mathbb{E}(\{s\},G)-\mathbb{E}(\{s\},G[V\setminus \{u\}])$.
\label{theo::block_expected}
\end{theorem}


{From Theorem~\ref{theo::block_expected}, it is evident that $\mathbb{E}[\sigma^{\rightarrow u}(s,g)]$ is the exact reduction in expected spread by blocking node $u$. Hence, we utilize the mean value of $\sigma^{\rightarrow u}(s,g)$ as an estimation for the decrease in expected spread for each node via random sampling. The subsequent theorem substantiates its efficacy as a reliable estimator for any given node $u$ and a set seed node $s$, provided an ample number of sampled graphs.
Let $\theta$ be the number of sampled graphs, $\xi^{\rightarrow u}(s,G)$ be the average number of $\sigma^{\rightarrow u}(s,g)$ and $OPT$ be the exact decrease of expected spread from blocking node $u$, i.e., $OPT=\mathbb{E}[\sigma^{\rightarrow u}(s,g)]$ (Theorem~\ref{theo::block_expected}).}


\begin{theorem}
\label{theorem:approx}
For seed node $s$ and a fixed node $u$, the inequality $|\xi^{\rightarrow u}(s,G)-OPT| < \varepsilon\cdot OPT$ holds with at least $1-{n^{-l}}$ probability when $\theta\ge \frac{l(2+\varepsilon)n\log n}{\varepsilon^2\cdot OPT}$.
\end{theorem}

{Theorem \ref{theorem:approx} proves that the mean value of $\sigma^{\rightarrow u}(s,g)$ provides a reliable estimate for the expected reduction in spread per node, which constitutes the essential theoretical accuracy assurance for subsequent sampling-based algorithms.}





\subsection{Dominator Trees of Sampled Graphs}
\label{sec::domi-tree}

In order to efficiently compute the decrease in expected spread resulting from the blocking of individual nodes, we consider constructing the dominator tree~\citep{AhoU73} for each sampled graph.
For the sampled graphs under the LT model, we can find that each node contains at most one incoming edge in each sampled graph, thus we first propose a linear algorithm for constructing dominator trees (Section~\ref{sec::linear-dominator}).
For the general graph, we apply the Lengauer-Tarjan algorithm \citep{LengauerT79} to construct the dominator tree (Section~\ref{sec::lt-algorithm}).


Note that, in the following subsection, the id of each node is reassigned by the sequence of a depth-first search ({DFS}) on the graph starting from the seed.

\begin{definition}[dominator]
Given $G=(V,E)$ and a source $s$, the node $u$ is a dominator of node $v$ when every path in $G$ from $s$ to $v$ has to go through $u$.
\end{definition}

\begin{definition}[immediate dominator]
Given $G=(V,E)$ and a source $s$, the node $u$ is the immediate dominator of node $v$, denoted $idom(v)=u$, if $u$ dominates $v$ and every other dominator of $v$ dominates $u$.
\end{definition}

We can find that every node except source $s$ has a unique immediate dominator. The dominator tree of graph $G$ is induced by the edge set $\{(idom(u),u)\mid u\in V\setminus \{s\}\}$ with root $s$ \citep{LowryM69}.
}

\noindent\textbf{\jiadong{Dominator Tree Construction under the LT Model.}}
\label{sec::linear-dominator}
\jiadong{
We can find that each node in the sampled graphs under the LT model has at most one incoming edge.
Thus, for each node in the sampled graphs, its immediate dominator is its unit in-neighbor.}
\jiadong{
Algorithm~A2 in the online appendix shows the details of constructing the structure of a dominator tree under the LT model, which can be implemented in $O(m)$ time.}


{\color{rehigh}\noindent\textbf{The Lengauer-Tarjan Algorithm.}
\label{sec::lt-algorithm}
It is an efficient algorithm for constructing the dominator tree.
It first computes the semidominator of each node $u$, denoted by $sdom(u)$, where $sdom(u)=\min \{v\mid \text{there is a path~} v=v_0,\cdots, v_k=u \text{~with~} v_i>u \text{~for any integer~} i\in [1, k)\}$. The semi-dominator can be computed by finding the minimum $sdom$ value on the paths of the DFS.
The details of the algorithm are in this paper~\citep{LengauerT79}.
The time complexity of the Lengauer-Tarjan algorithm is $O(m\cdot \alpha(m,n))$ which is almost linear, where $\alpha$ is the inverse function of Ackerman's function~\citep{ReelleZahlen}. 

\subsection{Estimation of Decrease in Expected Spread from Blocking Nodes or Edges}
\label{sec:new-es}

Following the above subsections, if node $u$ is blocked, we can use the number of nodes in the subtree rooted at $u$ in the dominator tree to estimate the decrease in expected spread. 
Thus, using DFS on dominator trees, we can estimate the decrease of expected spread for blocked nodes.

\begin{theorem}
Let $s$ be a fixed node in graph $g$. For any node $u\in (V(g)\setminus \{s\})$, we have $\sigma^{\rightarrow u}(s,g)$ equals to the size of the subtree rooted at $u$ in the dominator tree of graph $g$.
\label{theo:dominator-subtree}
\end{theorem}


Thus, for each blocker $u$, we can estimate the decrease of the expected spread by the average size of the subtrees rooted at $u$ in the dominator trees of the sampled graphs.

\begin{figure*}[!htbp]
\resizebox{0.85\linewidth}{!}{
\centering
\begin{subfigure}[h]{0.24\linewidth}
    \includegraphics[width=1\columnwidth]{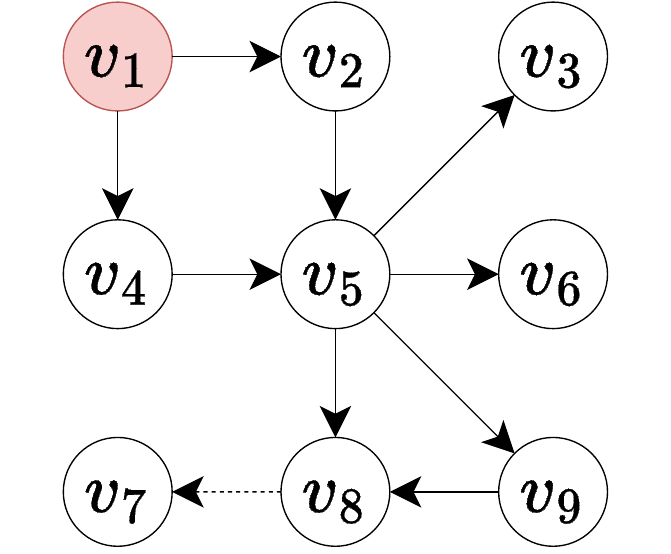}
    \caption{Sampled Graph 1}\label{fig:sample-example-a}
\end{subfigure}
\begin{subfigure}[h]{0.24\linewidth}
    \includegraphics[width=1\columnwidth]{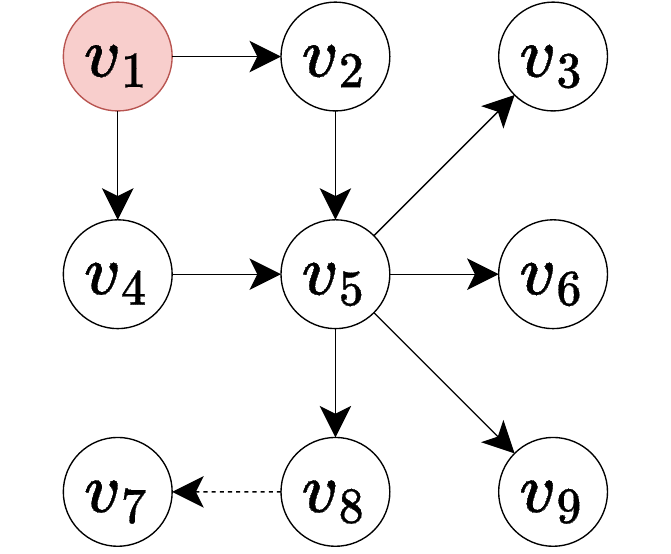}
    \caption{Sampled Graph 2}\label{fig:sample-example-b}
\end{subfigure}
\begin{subfigure}[h]{0.24\linewidth}
    \includegraphics[width=1\columnwidth]{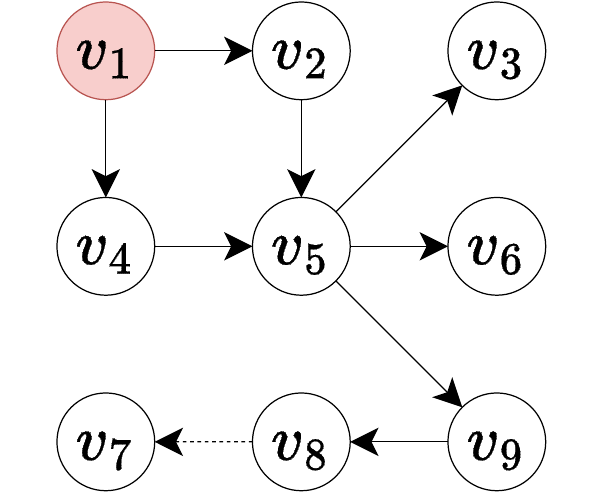}
    \caption{Sampled Graph 3}\label{fig:sample-example-c}
\end{subfigure}
\begin{subfigure}[h]{0.24\linewidth}
    \includegraphics[width=1\columnwidth]{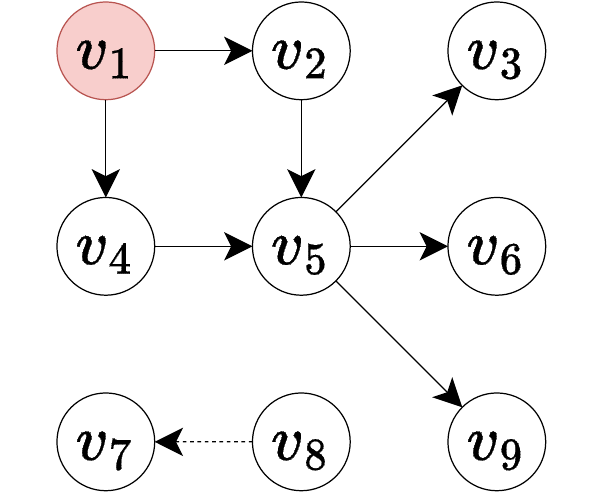}
    \caption{Sampled Graph 4}\label{fig:sample-example-d}
\end{subfigure}
}\centering
\caption{Sampled graphs of the graph $G$ in Figure~\ref{fig:sample}.}
\label{fig:sample-example}
\end{figure*}

\begin{figure*}[!htbp]
\centering
\resizebox{0.85\linewidth}{!}{
\centering
\begin{subfigure}[h]{0.245\linewidth}
    \includegraphics[width=1\columnwidth]{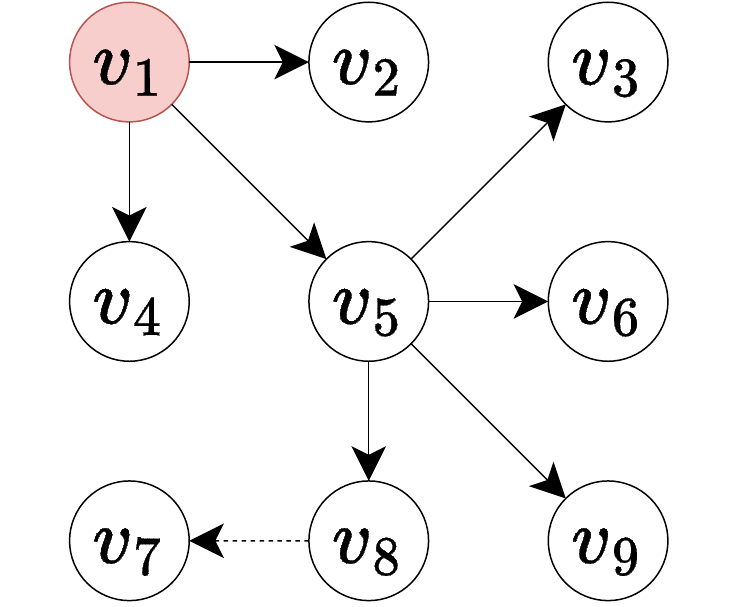}
    \caption{DT of Graph in Fig.~\ref{fig:sample-example-a}}\label{fig:sample-dt-example-a}
\end{subfigure}
\begin{subfigure}[h]{0.245\linewidth}
    \includegraphics[width=1\columnwidth]{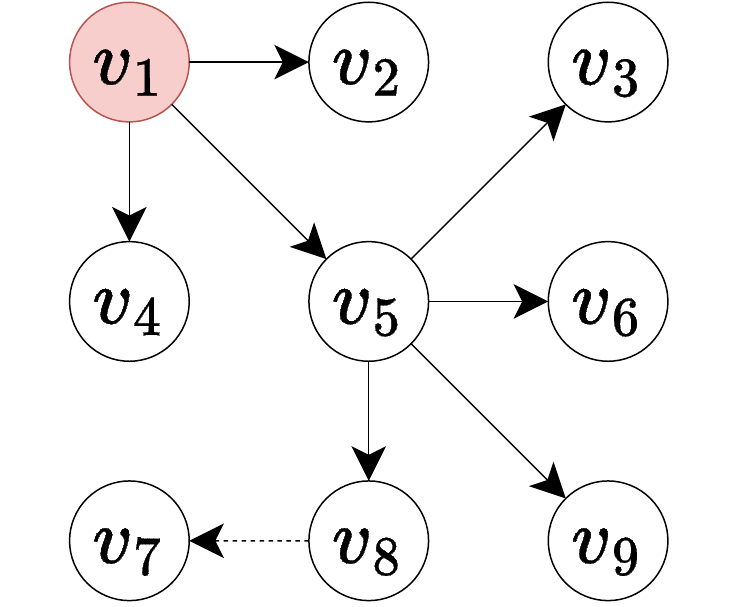}
    \caption{DT of Graph in Fig.~\ref{fig:sample-example-b}}\label{fig:sample-dt-example-b}
\end{subfigure}
\begin{subfigure}[h]{0.245\linewidth}
    \includegraphics[width=1\columnwidth]{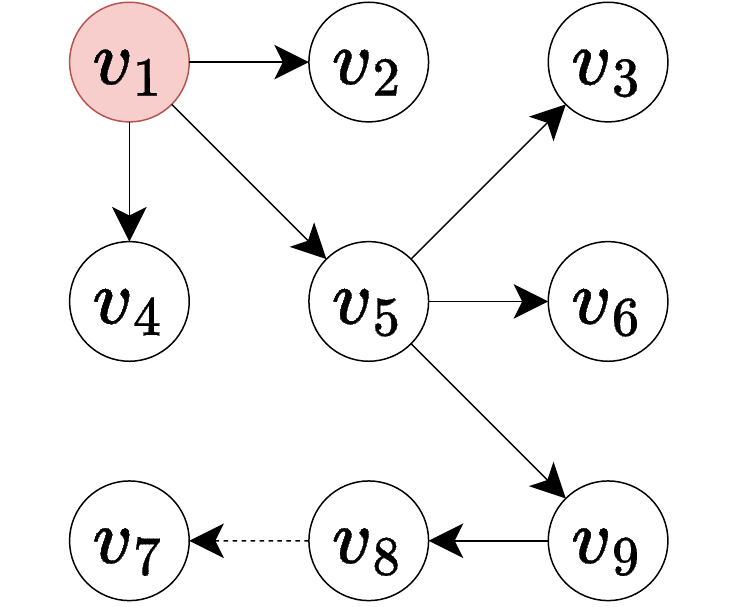}
    \caption{DT of Graph in Fig.~\ref{fig:sample-example-c}}\label{fig:sample-dt-example-c}
\end{subfigure}
\begin{subfigure}[h]{0.247\linewidth}
    \includegraphics[width=1\columnwidth]{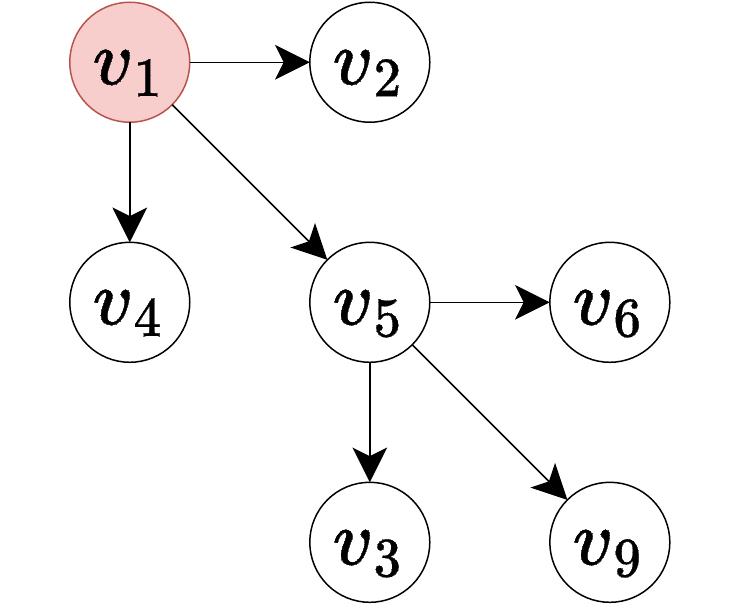}
    \caption{DT of Graph in Fig.~\ref{fig:sample-example-d}}\label{fig:sample-dt-example-d}
\end{subfigure}
}
\centering
\caption{Dominator trees of the sampled graphs in Figure~\ref{fig:sample-example}.}
\label{fig:sample-dt-example}
\end{figure*}

\begin{example}
{
\normalfont Given the graph $G$ illustrated in Figure~\ref{fig:sample}, there are only three edges with propagation probabilities less than $1$, specifically $(v_5,v_8)$, $(v_9,v_8)$ and $(v_8,v_7)$). The remaining edges are guaranteed to exist in any sampled graph.  
Figures~\ref{fig:sample-example-a}-\ref{fig:sample-example-d} display all possible sampled graphs. For brevity, the dotted edge $(v_8,v_7)$ indicates its potential presence or absence in a sampled graph, representing two distinct sampled graphs accordingly.
When $(v_8,v_7)$ is not in the sampled graphs, given $p_{v_5,v_8}=0.5$ and $p_{v_9,v_8}=0.2$, Figures~\ref{fig:sample-example-a}, \ref{fig:sample-example-b}, \ref{fig:sample-example-c} and \ref{fig:sample-example-d} have probabilities of existence equal to $0.1, 0.4, 0.1$ and $0.4$, respectively. 
With $p_{v_8,v_7}=0.1$, node $v_1$ can reach $8+0.1=8.1$ nodes in expectation in Figure~\ref{fig:sample-example-a}. Similarly, $v_1$ can reach $8.1,8.1$ and $7$ nodes (including $v_1$) in expectation in Figures~\ref{fig:sample-example-b}, \ref{fig:sample-example-c} and  \ref{fig:sample-example-d}, respectively.
Thus, the expected spread of graph $G$ is $7.66$, which is the same as the result in Example~\ref{example:compute}.}

Figures~\ref{fig:sample-dt-example-a}-\ref{fig:sample-dt-example-d} show the corresponding dominator trees of the sampled graphs in Figure~\ref{fig:sample-example}. For node $v_5$, the expected sizes of the subtrees rooted at $v_5$ are $5.1,5.1,5.1$ and $4$ in the dominator trees, respectively. Thus, the blocking of $v_5$ will lead to a $4.66$ decrease in the expected spread. As the sizes of subtrees of $v_2,v_3,v_4$, and $v_6$ are only $1$ in each dominator tree, blocking any of them will lead to a $1$ expected spread decrease. 
Similarly, blocking $v_7$, $v_8$ and $v_9$ will lead to $0.66$, $0.06$ and $1.11$ expected spread decrease, respectively.
\end{example}

\begin{algorithm}[!htbp]\color{rehigh}
    \SetVline 
    \SetFuncSty{textsf}
    \SetArgSty{textsf}
	\caption{DESC($G,s,\theta$)}
	\label{algo:blockes}
	\Input{a graph $G$, the source $s$ and the number of sampled graphs $\theta$}
	\Output{$\Delta [u]$ for each $u\in V(G)\setminus \{s\}$, i.e., the decrease of expected spread when $u$ is blocked}
	\State{$\Delta[\cdot]\leftarrow 0$}
	\For{$i\leftarrow 1$ to $\theta$}
	{
	    \State{Generate a sampled graph $g$ derived from $G$}
    	\State{$DT\leftarrow$ dominator tree of $g$ (Algorithm~A2 or Lengauer-Tarjan algorithm (Section~\ref{sec::lt-algorithm}))}
    	\State{$c[\cdot]\leftarrow$ the size of subtree in tree $DT$ when each node is the root}
            \State{$\Delta[u]\leftarrow \Delta[u] + c[u]/\theta$ \textbf{for} each $u\in V(g)\setminus \{s\}$}
	}

	\Return{$\Delta[\cdot]$}
\end{algorithm}
}

We propose the DESC algorithm for computing the decrease of the expected spread of each node, Algorithm~\ref{algo:blockes} shows the details. We set $\Delta[\cdot]$ as $0$ initially (Line 1). Then we generate $\theta$ different sampled graphs derived from $G$ (Lines 2-3). For each sampled graph, we first construct the dominator tree through Algorithm~A2 or Lengauer-Tarjan (Line 4). Then we use a simple DFS to compute the size of each subtree. After computing the average size of the subtrees and recording it in $\Delta[\cdot]$ (Line 6), we return $\Delta[\cdot]$ (Line 7).
As computing the sizes of the subtrees through DFS costs $O(m)$, Algorithm~\ref{algo:blockes} runs in $O(\theta\cdot m\cdot \alpha(m,n))$ under the IC model (resp. $O(\theta\cdot m)$ under the LT model).




\jiadong{For the IMIN-EB problem, we further need to estimate the decrease of expected spread when each edge is blocked in the graph.
We first construct edge-sampled graphs, denoted by $g^E$, based on graph $G$.
Let $g$ be a sampled graph based on graph $G$.
For each edge $(u,v)\in E(g)$, we use a virtual node $w_{u,v}$ to represent this edge. The edge-sampled graph $g^E$ contains two parts: $V_V$ and $V_E$. $V_V$ contains $n$ nodes where each node corresponds to $v\in V(g)$. $V_E$ contains $m$ nodes, i.e., $V_E=\cup_{(u,v)\in E(g)} w_{u,v}$. We then add edges $(u,w_{u,v})$ and $(w_{u,v},v)$ in the graph $g^E$ for each edge $(u,v)\in E(g)$. We can compute the decrease of expected spread led by edge $(u,v)$ through node $w_{u,v}$ on edge-sampled graphs.}

\vspace{-5mm}
\jiadong{
\begin{theorem}
Let $s$ be a fixed node, $g^E$ be a random edge-sampled graph derived from $G$ and $R_{u,v}$ be a set of nodes satisfying: (i) in the part $V_V$; (ii) reachable from $s$ in $g^E$ and (iii) all paths from $s$ in $g^E$ pass through $w_{u,v}$.
For any edge $(u,v)$, the decrease of expected spread by blocking $(u,v)$ is equal to $\mathbb{E}[|R_{u,v}|]$.
\label{theo::edge_expected}
\end{theorem}}


\jiadong{We then present how to compute the node set $R_{u,v}$ for each node $w_{u,v}$ in $V_E$.}

\vspace{-5mm}
\jiadong{\begin{theorem}
Let $s$ be a fixed node in graph $G$.
We construct a dominator tree of edge-sampled graph $g^E$ based on graph $G$.
For any node $w_{u,v}\in V_E$, we have $R_{u,v}$ are the nodes in $V_V$ and the subtree rooted at $w_{u,v}$ in the dominator tree of graph $g^E$.
\label{theo:edge-dominator-subtree}
\end{theorem}}


\begin{algorithm}[!htbp]
    \SetVline 
    \SetFuncSty{textsf}
    \SetArgSty{textsf}
	\caption{\jiadong{DESCE($G,s,\theta$)}}
	\label{algo:edge-estim}
	\Input{\jiadong{a graph $G$, the source $s$ and the number of sampled graphs $\theta$}}
	\Output{\jiadong{$\Delta [u,v]$, i.e., the decrease of expected spread when $(u,v)$ is blocked}}
	\State{\jiadong{$\Delta[\cdot]\leftarrow 0$}}
	\For{\jiadong{$i\leftarrow 1$ to $\theta$}}
	{
	    \State{\jiadong{Generate a sampled graph $g$ derived from $G$}}
            \State{\jiadong{$V_V\leftarrow V(g)$}}
            \State{\jiadong{Generate a edge-sampled graph $g^E$ based on $g$}}
    	\State{\jiadong{Construct a dominator tree $DT$ of graph $g^E$}}
    	\State{\jiadong{$R[\cdot]\leftarrow V_V\cap $ the nodes in subtree in tree $DT$ when each node is the root}}
            \State{\jiadong{$\Delta[u,v]\leftarrow \Delta[u,v] + |R[u,v]|/\theta$ \textbf{for} each $w_{u,v}\in V(g^E)\setminus V_V$}}
	}

	\Return{\jiadong{$\Delta[\cdot]$}}
\end{algorithm}

\jiadong{Based on Theorem~\ref{theo::edge_expected} and~\ref{theo:edge-dominator-subtree}, we propose the DESCE algorithm (Algorithm~\ref{algo:edge-estim}) for computing the decrease of the expected spread of each edge. We set $\Delta[\cdot]$ as $0$ initially (Line 1). Then we generate $\theta$ different edge-sampled graphs derived from $G$ (Lines 2-5). For each sampled graph, we first construct the dominator tree through Algorithm~A2 or Lengauer-Tarjan (Line 6). Then we can use a DFS to compute the size of $R_{u,v}$ from leaves to root (Line 7). After computing the average size of the $R_{u,v}$ and recording it in $\Delta[\cdot]$ (Line 8), we return $\Delta[\cdot]$ (Line 9).}
\jiadong{As computing the sizes of the $R_{u,v}$ through DFS costs $O(m)$, Algorithm~\ref{algo:edge-estim} runs in $O(\theta\cdot m\cdot \alpha(m,n))$ under the IC model (resp. $O(\theta\cdot m)$ under the LT model).}

%% file: imin.tex
\section{Our Proposed Algorithms}
\label{sec::app-IMIN}

{In this section, applying the new framework of expected spread estimation for selecting the candidates (Algorithm~\ref{algo:blockes}), we propose our AdvancedGreedy algorithm (Section~\ref{sec:advanced-greedy}) with higher efficiency without sacrificing the effectiveness, compared with the baseline.
We then prove our AdvancedGreedy algorithm can achieve a $(1-1/e-\varepsilon)$-approximation under the LT model. However, the problems are hard to approximate under the IC model.
As the greedy approaches do not consider the cooperation of candidate blockers during the selection, some important nodes may be missed.
Thus, we further propose a superior heuristic, the GreedyReplace algorithm, to achieve a better result quality under the IC model (Section~\ref{sec:greedyreplace}).}

\subsection{The AdvancedGreedy Algorithm}
\label{sec:advanced-greedy}

Based on Sections~\ref{sec:baseline} and~\ref{sec:new-es}, we propose AdvancedGreedy algorithm with high efficiency.
In the algorithm, we aim to greedily find the node $u$ (resp. edge $(u,v)$) that leads to the largest decrease in the expected spread. Algorithm~\ref{algo:blockes} and \ref{algo:edge-estim} can efficiently compute the expected spread decrease of every candidate node/edge.
Thus, we can directly choose the node/edge which can cause the maximum decrease of expected spread to block.

\begin{algorithm}[!htbp]
    \SetVline 
    \SetFuncSty{textsf}
    \SetArgSty{textsf}
	\caption{AdvancedGreedy($G,s,b,\theta,CB$)}
	\label{algo:fast-greedy}
	\Input{a graph $G$, seed $s$, budget $b$, the number of sampled graphs $\theta$, and candidate node/edge set $CB$}
	\Output{the blocked node/edge set $B$}
    \State{$B\leftarrow$ empty}
	\For{$i\leftarrow 1$ to $b$}
	{
	    \State{$\Delta[\cdot] \leftarrow$ DESC($G[V\setminus B],s,\theta$) or DESCE($G[V\setminus B],s,\theta$)}
	    \State{$B\leftarrow B\cup \{x\}$, where $x=  \argmax_{c\in CB} \Delta[c]$}
	}
	\Return{$B$}
\end{algorithm}

Algorithm~\ref{algo:fast-greedy} presents the details of our AdvancedGreedy algorithm.
For the IMIN-EB problem, we cannot block all nodes in the sampled graph (i.e., can block $V_E$ in the edge-sampled graph $g^E$), thus we need a candidate blocker set $CB$ as the input of the algorithm. For IMIN problem, $CB=V(G)\setminus \{s\}$, but for IMIN-EB problem, $CB=V_E=\cup _{(u,v)\in E(G)}w_{u,v}$.
We start with the empty blocker set (Line 1). In each of the $b$ iterations (Line 2), we first estimate the decrease of the expected spread of each node (Line 3), find the node $x$ such that $\Delta[x]$ is the largest as the blocker and insert it to blocker set (Line 4). Finally, the algorithm returns the blocker set $B$ (Line 5).

\noindent \textbf{Comparison with Baseline Algorithm.}
One round of MCS on $G$ generates a graph $G'$ where $V(G')=V(G)$ and each edge in $E(G)$ appears in $G'$ if the simulation picks this edge.
Thus, if we have $r=\theta$, our computation based on sampled graphs will not sacrifice the effectiveness, compared with MCS.
For efficiency, Algorithm~\ref{algo:fast-greedy} runs in $O(b\cdot \theta \cdot m\cdot \alpha(m,n) )$ for both two problems and the time complexity of the baseline (Algorithm~A1) are $O(b\cdot r\cdot m\cdot n)$ and $O(b\cdot r\cdot m\cdot m)$. As $\alpha(m,n)$ is much smaller than $n$ and $m$, our AdvancedGreedy algorithm has a significantly lower time complexity, compared with the baseline algorithm.

\noindent \textbf{Approximation Guarantee of the AdvancedGreedy Algorithm.}
In the IC model, we establish that the expected spread function is monotonic and not supermodular for both the IMIN and IMIN-EB problems (Theorem~8). Then we show that the IMIN problem is APX-hard unless P=NP when $\frac{n+m-b}{OPT}=c$, where $OPT$ represents the optimal solution for the IMIN problem (Theorem~9).
Under the LT model, we find that the expected spread function is both monotonic and supermodular for both problems as outlined in Theorems~10 and 12.
Given that the expected spread is non-increasing and supermodular with respect to the blocker set $B$, the AdvancedGreedy algorithm provides a $(1-1/e-\varepsilon)$-approximation to the optimal solution for both the IMIN and IMIN-EB problems under the LT model~\citep{tzoumas2016sensor}.

{\color{rehigh}\subsection{The GreedyReplace Algorithm}
\label{sec:greedyreplace}

As the greedy algorithm does not achieve an approximate guarantee under the IC model, we further propose the GreedyReplace algorithm to improve its effectiveness.
We can find that some out-neighbors of the seed may be an essential part of the result while they may be missed by current greedy heuristics. Thus, we propose a new heuristic (GreedyReplace) which is to first select $b$ out-neighbors of the seed as the initial blockers, and then greedily replace a blocker with another node if the expected spread will decrease.

\begin{example}
\normalfont Considering the graph in Figure~\ref{fig:sample} with the seed $v_1$.
When $b=1$, Greedy chooses $v_5$ as the blocker because it leads to the largest expected spread decrease ($v_3,v_6,v_7,v_8$, and $v_9$ will not be influenced by $v_1$, i.e., the expected spread is $3$).
When $b = 2$, it further blocks $v_2$ or $v_4$ in the second round, and the expected spread becomes $2$.
OutNeighbors only considers blocking $v_2$ and $v_4$. It blocks either of them when $b=1$ ($\mathbb{E}(\cdot)=6.66$), and blocks both of them when $b=2$ ($\mathbb{E}(\cdot)=1$).
\end{example}

	


In this example, we find that the performance of Greedy algorithm is better than the OutNeighbors when $b=1$, but its expected spread may become larger than OutNeighbors with the increase of $b$. 
As budget $b$ can be either small or large in different applications, it is essential to further improve the heuristic algorithm.

Due to the above motivation, we propose the GreedyReplace algorithm as a solution to address the defects of the Greedy and OutNeighbors algorithms, while also combining their advantages.
We first greedily choose $b$ out-neighbors of the seed as the initial blockers. 
Then, we replace the blockers according to the reverse order of the out-neighbors' blocking order. 
As we can use Algorithm~\ref{algo:blockes} to compute the decrease of the expected spread of blocking any other node, in each round of replacement, we set all the nodes in $V(G)\setminus (\{s\}\cup B)$ as the candidates for replacement.
We will early terminate the replacement procedure when the node to replace is the current best blocker.

The expected spread of GreedyReplace is certainly not larger than the algorithm which only blocks the out-neighbors. Through the trade-off between choosing the out-neighbors and the replacement, the cooperation of the blockers is considered in GreedyReplace. 

\begin{example}
\label{example:move}
\normalfont Considering the graph in Figure~\ref{fig:sample} with the seed $v_1$.
When $b=1$, GreedyReplace first considers the out-neighbors as the candidate blockers and sets $v_2$ or $v_4$ as the blocker. As blocking $v_5$ can achieve a smaller influence spread than both $v_2$ and $v_4$, it will replace the blocker with $v_5$, thus the expected spread is $3$. When $b=2$, GreedyReplace first block $v_2$ and $v_4$, and there is no better node to replace. The expected spread is $1$. GreedyReplace achieves the best performance for either $b=1$ or $b=2$, compared to the Greedy and OutNeighbors algorithms. 
\end{example}

\begin{algorithm}[!htbp]\color{rehigh}
    \SetVline 
    \SetFuncSty{textsf}
    \SetArgSty{textsf}
	\caption{GreedyReplace($G,s,b,\theta$)}
	\label{algo:greedyreplace}
	\Input{a graph $G$, the source $s$, budget $b$ and the number of sampled graphs $\theta$}
	\Output{the blocker set $B$}
	\State{$CB\leftarrow N^{out}_s$; $B\leftarrow$ empty}
	\For{$i\leftarrow 1$ to $\min\{d^{out}_s,b\}$}
	{
	    \State{$\Delta[\cdot] \leftarrow$ DESC($G[V\setminus B],s,\theta$)}
	    \State{$x\leftarrow  \argmax_{u\in CB}\Delta[u]$; $CB\leftarrow CB\setminus \{x\}$; $B\leftarrow B\cup \{x\}$}
	}
	
	\For{each $u\in B$ with the reversing order of insertion}
	{
	    \State{$B\leftarrow B\setminus \{u\}$; $\Delta[\cdot] \leftarrow$ DESC($G[V\setminus B],s,\theta$)}
	    \State{$x\leftarrow  \argmax_{u\in V(G)\setminus B}\Delta[u]$; $B\leftarrow B\cup \{x\}$}
            \State{\textbf{if} $u=x$ \textbf{then} Break}
	}
	\Return{$B$}
\end{algorithm}


Algorithm~\ref{algo:greedyreplace} shows the details of GreedyReplace. We first push all out-neighbors of the seed into candidate blocker set $CB$ and set blocker set empty (Line 1). For each round, we choose the candidate blocker which leads to the largest expected spread decrease as the blocker, and then update $CB$ and $B$ (Lines 3-4). 
Then we consider replacing the blockers in $B$ by the reversing order of their insertions (Line 5). We remove the replaced node from the blocker set and use Algorithm~\ref{algo:blockes} to compute the decrease of expected spread $\Delta[\cdot]$ for each candidate blocker (Line 6).
We use $x$ to record the node with the largest spread decrease computed so far, by enumerating each of the candidate blockers (Line 7). 
If the node to replace is the current best blocker, we will early terminate the replacement (Line 8). Algorithm~\ref{algo:greedyreplace} returns the set $B$ of $b$ blockers (Line 9).

The time complexity of GreedyReplace is $O(\min \{d^{out}_s,b\}\cdot \theta \cdot m\cdot \alpha(m,n))$. As its time complexity hinges on the edge count in the sampled graphs, in practice the time cost is much less than the worst case.}

%% file: MS_Version/exp_MS.tex
\section{Experiments}
\label{sec:exp}


        


\noindent\textbf{Datasets.}
\jiadong{The experiments are conducted on $8$ real-world datasets, which are all publicly available from SNAP ({\url{http://snap.stanford.edu}}).
Table~\ref{tab:dat} shows the statistics of the datasets, ordered by the number of edges in each dataset, where $d_{avg}$ is the average node degree (the sum of in-degree and out-degree for each directed graph) and $d_{max}$ is the largest node degree.
For an undirected graph, we consider each edge as bi-directional. These experiments cover a diverse range of networks, spanning from networks consisting of 1,000 nodes and 25,000 edges to networks comprising 1 million nodes and 3 million edges.}

\noindent \textbf{Propagation Models.} 
Our algorithm is designed for the IC and LT models with nonuniform propagation probabilities $p_{u,v}$ for each directed edge $(u,v)$. To generate these nonuniform probabilities $p_{u,v}$, we employ two propagation models following the existing literature:
(i). Trivalency (TR) model \citep{JungHC12,maximization1,min-greedy-tree}: The propagation probability $p_{u,v}$ for each edge $(u,v)$ is randomly chosen from the set \{0.1, 0.01, 0.001\}, which correspond to high, medium, and low influences, respectively.
(ii). Weighted cascade (WC) model \citep{first-max,max-rr}: The propagation probability $p_{u,v}$ for each edge $(u,v)$ is $1/d^{in}_v$, where $d^{in}_v$ is the in-degree of $v$.

\vspace{-3mm}
\begin{table}[!htbp]\color{jiadong}
	\caption{\jiadong{Statistics of Datasets}}
	\centering%
	\renewcommand{\arraystretch}{1.3}%
	\resizebox{\linewidth}{!}{
	\begin{tabular}{|lc|lcl|ccccc|}
	\toprule
	\bf{Dataset} & Abbr. & Description & Node & Edge $(u,v)$ & \bf{$n$} & \bf{$m$} & \bf{$d_{avg}$} & \bf{$d_{max}$} & \bf{Type}  \\
		\midrule
		EmailCore & EC & email data& member & $u$ sent at least one email to $v$ & 1,005 & 25,571 &  49.6 &544 & Directed\\
		Facebook & F & social network & user & $u$ and $v$ are friends & 4,039 & 88,234 & 43.7 & 1,045  & Undirected \\
		Wiki-Vote & W &vote data & user &  $u$ voted on user $v$  & 7,115 & 103,689 & 29.1 & 1,167 & Directed\\

		EmailAll & EA& email data& user & $u$ sent at least one email to $v$  & 265,214 & 420,045 & 3.2 & 7,636 & Directed\\
		DBLP & D&bibliography & author & $u$ and $v$ are co-authors & 317,080 & 1,049,866 & 6.6 & 343 & Undirected\\
		
		Twitter & T &social network & user & $u$ and $v$ are friends & 81,306 & 1,768,149 & 59.5& 10,336  & Directed\\
		
		Stanford & S & website pages & page & $u$ has hyperlinks to $v$ & 281,903 & 2,312,497 & 16.4 & 38,626 & Directed\\
		Youtube & Y&social network & user & $u$ and $v$ are friends  & 1,134,890 & 2,987,624 & 5.3 & 28,754 & Undirected\\
		
		\bottomrule
	\end{tabular}
	}
	\label{tab:dat}
\end{table}

\noindent \textbf{Settings.}
Unless otherwise specified, we set $r=10^4$ for MCS
, and for our sampled-graph-based algorithms (AdvancedGreedy and GreedyReplace), we sample $10^4$ graphs in the experiments. In each of our experiments, we independently execute each algorithm 5 times and report the average result. By default, we terminate an algorithm if the running time reaches 24 hours.
Based on the results of the $\theta$ selection study (details in the online appendix 4), we set $\theta=10^4$ for all experiments.

\noindent \textbf{Environments.} The experiments are performed on a CentOS Linux server (Release 7.5.1804) with Quad-Core Intel Xeon CPU (E5-2640 v4 @ 2.20GHz) and 128G memory. All the algorithms are implemented in C++. The source code is compiled by GCC(7.3.0) under O3 optimization.

\noindent \textbf{Algorithms.}
The experiments involve comparing our proposed algorithms, namely the GreedyReplace and AdvancedGreedy, with both the greedy algorithms and other basic heuristics mentioned in the existing literature. The algorithms evaluated in our experiments are listed below.

\noindent 1. \underline{\texttt{Exact}}
: This algorithm identifies the optimal solution by searching all possible combinations of $b$ blockers and uses MCS with $r=10^4$ to compute the expected spread of candidate set of blockers. The algorithms show optimal effectiveness without considering efficiency, providing a benchmark for effectiveness comparison.

\noindent 2. \underline{\texttt{Rand} (RA)}: This algorithm randomly chooses $b$ blockers (excluding the seeds) in the graph.

\noindent 3. \underline{\texttt{OutDegree} (OD)} \citep{Nature-error,Viruses}: This algorithm selects $b$ nodes with the highest out-degrees as the blockers for the IMIN problem and selects $b$ directed edges where the ending nodes have the highest out-degrees {for IMIN-EB problem}.

{\noindent 4. \underline{\texttt{GRASP} (GS)} \citep{FeoR95}: Greedy Randomized Adaptive Search (GRASP) is a metaheuristic algorithm commonly utilized for combinatorial optimization problems. It involves iterations comprising the sequential construction of a greedy randomized solution followed by iterative enhancements through local search. We run it one iteration since it needs a large time cost.}

\noindent 5. \underline{\texttt{BaselineGreedy} (BG)} \citep{min-greedy2,min-greedy1,min-greedy-tree}: This algorithm represents the current state-of-the-art approach. We have introduced and discussed in detail in Section \ref{sec:baseline}. It leverages the MCS technique to estimate the expected spread.

\noindent 6. \underline{\texttt{AdvancedGreedy} (AG)}: Our proposed AdvancedGreedy algorithm (Algorithm~\ref{algo:fast-greedy}), which applies dominator tree techniques for estimating the reduction of expected spread.

\noindent 7. \underline{\texttt{GreedyReplace} (GR)}: Our proposed GreedyReplace algorithm (Algorithm~\ref{algo:greedyreplace}). It enhances the AdvancedGreedy algorithm by considering the relationships among candidate blockers.

{\noindent \textbf{Availability.}
The datasets, source code, and detailed results are available online~\citep{IM2024}.}

\noindent \textbf{Comparison with the Exact Algorithm.}
\jiadong{As AG has an approximate guarantee under the LT model in both IMIN problem and IMIN-EB problem, we only compare the result of GR with the \texttt{Exact} algorithm under the IC model in two problems.
Due to the huge time cost of \texttt{Exact}, we extract small datasets by iteratively extracting a node and all its neighbors, until the number of extracted nodes reaches $100$ in IMIN problem (resp. edges reaches $100$ in IMIN-EB problem). 
For \texttt{EmailCore}, we extract $5$ such subgraphs.
We randomly choose $10$ nodes as the seeds. As the graph is small, we can use the exact computation of the expected spread~\citep{MaeharaSI17} for comparison between \texttt{Exact} and GR.
We demonstrate that the expected spread of the GR algorithm closely aligns with the results of the Exact algorithm in both the TR Model (Tables~\ref{tab:exact} Panel A and Panel B) and the WC Model (Table A2 Panel C and Panel D in online appendix). Additionally, the GR algorithm outperforms the Exact algorithm in terms of speed, with a difference of up to six orders of magnitude in both the IMIN problem and the IMIN-EB problem. The results indicate that our proposed algorithms are highly effective and can achieve optimal results in a shorter timeframe.}

\begin{table}[!htbp]\color{jiadong}
    \centering
\caption{Exact v.s. GreedyReplace (IC model)}\label{tab:exact}
\resizebox{0.6\linewidth}{!}{
\begin{tabular}{|c|c|c|c|c|c|c|c|c|c|c|}
\toprule
\multirow{2}{*}{ b } & \multicolumn{3}{c|}{Expected Spread} & \multicolumn{2}{c|}{Running Time (s)} & \multicolumn{3}{c|}{Expected Spread} & \multicolumn{2}{c|}{Running Time (s)}  \\ \cline{2-11}
& Exact & GR & Ratio & Exact & GR & Exact & GR & Ratio & Exact & GR \\ 
\midrule
& \multicolumn{5}{c|}{Panel A: IMIN Problem (TR Model)} & \multicolumn{5}{c|}{Panel B: IMIN-EB Problem (TR Model)}\\
\midrule
1 & 12.614 & 12.614 & 100\% & 3.07 & 0.12 & 10.873 & 10.873 & 100\% & 3.04 & 0.10 \\
2 & 12.328 & 12.334 & 99.95\% & 130.91 & 0.21 & 10.652 & 10.655 & 99.9\% & 150.72 & 0.31 \\
3 & 12.112 & 12.119 & 99.94\% & 3828.2 & 0.25 & 10.454 & 10.459 & 99.9\% & 5373.9 & 0.35 \\
4 & 11.889 & 11.903 & 99.88\% & 80050 & 0.33 & 10.360 & 10.407 & 99.5\% & 90897 & 0.40 \\ 
\midrule
\bottomrule
\end{tabular}
}
\vspace{-5mm}
\end{table}

\begin{table*}[!htbp]
\centering
\caption{Expected Spread in the IMIN problem (IC model)}\label{table:heuristics}
\resizebox{\linewidth}{!}{
\begin{tabular}{c|lllll|lllll|lllll|lllll}
\toprule
\multirow{2}{*}{\centering b } & \multicolumn{5}{c|}{EmailCore (TR model)} & \multicolumn{5}{c|}{Facebook (TR model)} & \multicolumn{5}{c|}{Wiki-Vote (TR model)} & \multicolumn{5}{c}{EmailAll (TR model)} \\
& RA & OD & GS & AG & GR & RA & OD & GS & AG & GR & RA & OD & GS & AG & GR & RA & OD & GS & AG & GR \\
\midrule
20 & 354.88$^{***}$ & 230.10$^{***}$ & 223.45$^{**}$ & 220.59$^{**}$ & \bf{219.69}
& 16.059$^{***}$ & 16.026$^{***}$&12.134$^{**}$ & 11.717$^{*}$ & \bf{11.691}
& 512.62$^{***}$ & 325.51$^{***}$&135.935$^{***}$ & 131.30$^{*}$ & \bf{130.77}
& 548.99$^{***}$ & 286.05$^{***}$ &14.925$^{***}$& 14.642$^{**}$ & \bf{13.640} \\
40 & 341.33$^{***}$ & 98.712$^{***}$ &85.193$^{***}$& 84.022$^{*}$ & \bf{83.823}
& 16.037$^{***}$ & 16.019$^{***}$ &10.509$^{*}$ & 10.416$^{*}$ & \bf{10.413}
& 512.18$^{***}$ & 222.00$^{***}$ &45.931$^{***}$ & 46.747$^{***}$ & \bf{43.898}
& 546.94$^{***}$ & 221.97$^{***}$ &10.310$^{*}$ & 10.319$^{*}$ & \bf{10.002} \\
60 & 325.13$^{***}$ & 47.249$^{***}$ & 35.092$^{**}$ & 35.085$^{**}$ & \bf{33.634}
& 16.033$^{***}$ & 16.010$^{***}$ &10.151$^{*}$& 10.151$^{*}$ & \bf{10.149}
& 507.11$^{***}$ & 138.60$^{***}$ &25.013$^{***}$ & 25.514$^{***}$ & \bf{23.282}
& 546.39$^{***}$ & 148.52$^{***}$ &10.003$^{*}$& 10 & 10 \\
80 & 304.90$^{***}$ & 30.277$^{***}$ &18.996$^{*}$ & 19.001$^{*}$ & \bf{18.848}
& 15.997$^{***}$ & 15.987$^{***}$ &10.031$^{*}$& 10.028$^{*}$ & \bf{10.026}
& 501.49$^{***}$ & 32.646$^{***}$ &17.398$^{*}$ & 17.332$^{*}$ & \bf{17.322}
& 545.41$^{***}$ & 100.84$^{***}$ &10& 10 & 10 \\
100 & 283.54$^{***}$ & 22.696$^{***}$ &13.595$^{*}$ & 13.640$^{*}$
& \bf{13.533} & 15.994$^{***}$ & 15.980$^{***}$&10.003$^{*}$ & 10.001 & 10.001
& 496.05$^{***}$ & 25.831$^{***}$ &14.720$^{*}$ & 14.726$^{*}$ & \bf{14.518}
& 544.59$^{***}$ & 55.398$^{***}$ &10& 10 & 10 \\
\bottomrule

\toprule
\multirow{2}{*}{\centering b } & \multicolumn{5}{c|}{DBLP (TR model)} & \multicolumn{5}{c|}{Twitter (TR model)} & \multicolumn{5}{c|}{Stanford (TR model)} & \multicolumn{5}{c}{Youtube (TR model)} \\
& RA & OD & GS & AG & GR & RA & OD & GS & AG & GR & RA & OD & GS & AG & GR & RA & OD & GS & AG & GR \\
\midrule
20 & 13.747$^{***}$ & 13.730$^{***}$ & 10.508$^{*}$& 10.502$^{*}$ & \bf{10.499}
& 16,801$^{***}$ & 16,610$^{***}$ & 16,103$^{**}$ &16,101$^{*}$ & \bf{16,100}
& 16.087$^{***}$ & 16.075$^{***}$ & 11.872$^{***}$ &12.069$^{***}$ & \bf{10.483}
& 14.774$^{***}$ & 14.762$^{***}$ &13.431$^{***}$ & 14.743$^{***}$ & \bf{10.950} \\

40 & 13.739$^{***}$ & 13.725$^{***}$ & 10.084$^{*}$& 10.079 & 10.079
& 16,796$^{***}$ & 16,470$^{***}$ & 15,749$^{**}$ & 15,749$^{**}$ & \bf{15,748}
& 16.080$^{***}$ & 16.071$^{***}$ & 10.493$^{*}$ & 10.488$^{*}$ & \bf{10.234}
& 14.773$^{***}$ & 14.755$^{***}$ & 10.009$^{*}$ &10.075$^{*}$ & \bf{10.002} \\

60 & 13.737$^{***}$ & 13.721$^{***}$ & 10.014$^{*}$& 10.012$^{*}$ & \bf{10.010}
& 16,786$^{***}$ & 16,329$^{***}$ & 15,015$^{***}$ &15,447$^{***}$ & \bf{14,972}
& 16.071$^{***}$ & 16.040$^{***}$ & 10.115$^{*}$ & 10.136$^{*}$ & \bf{10.075}
& 14.773$^{***}$ & 14.750$^{***}$ & 10& 10 & 10 \\

80 & 13.720$^{***}$ & 13.714$^{***}$ & 10 & 10 & 10
& 16,780$^{***}$ & 16,175$^{***}$ & 14,524$^{***}$ &14,610$^{***}$ & \bf{14,474}
& 16.064$^{***}$ & 16.017$^{***}$ & 10.022$^{*}$ &10.026$^{*}$ & \bf{10.019}
& 14.767$^{***}$ & 14.742$^{***}$ & 10& 10 & 10 \\

100 & 13.716$^{***}$ & 13.706$^{***}$& 10  & 10 & 10 &
16,771$^{***}$ & 16,057$^{***}$ & 13,439$^{***}$ &13,619$^{***}$ & \bf{13,181}
& 16.052$^{***}$ & 15.989$^{***}$ & 10.008$^{*}$ &10.009$^{*}$ & \bf{10.002}
& 14.762$^{***}$ & 14.729$^{***}$ & 10& 10 & 10 \\
\bottomrule

\multicolumn{13}{l}{\begin{minipage}{1.1\textwidth}  Notes. Best performers are bold. All experiments are repeated five times.

~~~~~~~~~~~~~~$^{*}p<0.1$; $^{**}p<0.05$; $^{***}p<0.001$.
\end{minipage}}

\end{tabular} }
\vspace{-5mm}
\end{table*}

\begin{table*}[!htbp]
\vspace{-4mm}
\centering
\caption{\jiadong{Expected Spread in the IMIN-EB problem (IC model)}}\label{table:heuristics-eb}
\resizebox{\linewidth}{!}{
\begin{tabular}{c|lllll|lllll|lllll|lllll}
\toprule
\multirow{2}{*}{\centering b } & \multicolumn{5}{c|}{EmailCore (TR model)} & \multicolumn{5}{c|}{Facebook (TR model)} & \multicolumn{5}{c|}{Wiki-Vote (TR model)} & \multicolumn{5}{c}{EmailAll (TR model)} \\
& RA & OD & GS & AG & GR & RA & OD & GS & AG & GR & RA & OD & GS & AG & GR & RA & OD & GS & AG & GR \\
\midrule
20 & 364.62$^{***}$ & 364.42$^{***}$ & 362.652$^{**}$& 361.54$^{*}$ & \bf{361.51}
& 57.846$^{***}$ & 58.04$^{***}$ & 33.349$^{*}$ &33.346$^{*}$ & \bf{33.279}
& 484.60$^{***}$ & 487.94$^{***}$ & 167.17$^{**}$ & 167.93$^{**}$ & \bf{166.79}
& 1,567.0$^{***}$ & 1566.2$^{***}$ & 354.67$^{***}$ & 361.51$^{***}$ & \bf{338.46} \\
40 & 364.53$^{***}$ & 364.36$^{***}$ & 356.98$^{*}$ &357.15$^{*}$ & \bf{356.48}
& 57.829$^{***}$ & 57.829$^{***}$ & 23.751$^{*}$ &23.748$^{*}$ & \bf{23.702}
& 481.22$^{***}$ & 482.27$^{***}$ & 41.025$^{**}$ &42.768$^{**}$ & \bf{40.205}
& 1,561.9$^{***}$ & 1564.1 $^{***}$& 82.792$^{***}$ & 83.462$^{***}$ & \bf{81.488} \\
60 & 364.42$^{***}$ & 364.23$^{***}$ & 335.16$^{***}$ &336.62$^{***}$ & \bf{333.41}
& 57.669$^{***}$ & 57.579$^{***}$ & 18.907$^{*}$ & 18.901$^{*}$ & \bf{18.859}
& 480.61$^{***}$ & 482.20$^{***}$ & 16.991$^{*}$ &17.329$^{**}$ & \bf{16.311}
& 1,556.6$^{***}$ & 1561.0$^{***}$ & 25.204$^{**}$& 28.143$^{***}$ & \bf{23.414} \\
80 & 364.30$^{***}$ & 364.10$^{***}$ & 275.15$^{***}$ &281.97$^{***}$ & \bf{267.79}
& 57.568$^{***}$ & 57.363$^{***}$ & 15.631$^{*}$ &15.624$^{*}$ & \bf{15.611}
& 479.28$^{***}$ & 475.20$^{***}$ & 11.489$^{*}$ &11.491$^{*}$ & \bf{11.360}
& 1,556.6$^{***}$ & 1558.2$^{***}$ & 13.049$^{**}$ & 14.049$^{**}$ & \bf{12.954} \\
100 & 364.12$^{***}$ & 363.81$^{***}$ & 197.25$^{***}$ & 206.65$^{***}$ & \bf{189.50}
& 57.322$^{***}$ & 57.316$^{***}$ & 13.329 & 13.330$^{*}$ & {13.329}
& 475.67$^{***}$ & 472.55$^{***}$ & 10.097$^{*}$ &10.101$^{*}$ & \bf{10.086}
& 1,551.9$^{***}$ & 1550.7$^{***}$ & 11.628$^{*}$ & 11.783$^{*}$ & \bf{11.590} \\
\bottomrule
\toprule
\multirow{2}{*}{\centering b } & \multicolumn{5}{c|}{DBLP (TR model)} & \multicolumn{5}{c|}{Twitter (TR model)} & \multicolumn{5}{c|}{Stanford (TR model)} & \multicolumn{5}{c}{Youtube (TR model)} \\
& RA & OD&GS & AG & GR & RA & OD&GS & AG & GR & RA & OD&GS & AG & GR & RA & OD&GS & AG & GR \\
\midrule
20 & 10.690$^{**}$ & 10.692$^{***}$ & 10.448$^{*}$& 10.451$^{*}$ & \bf{10.354}
& 16,800$^{***}$ & 16,800$^{***}$ & 16,703$^{**}$ & 16,701$^{**}$ & \bf{16,697}
& 13.477$^{***}$ & 13.456$^{***}$ & 11.294$^{***}$ & 11.540$^{***}$ & \bf{10.989}
& 10.819$^{**}$ & 10.817$^{**}$ & 10.079$^{*}$ & 10.081$^{*}$ & \bf{10.069} \\

40 & 10.686$^{**}$ & 10.688$^{**}$ & 10.251$^{*}$ & 10.253$^{*}$ & \bf{10.144}
& 16,800$^{***}$ & 16,799$^{***}$ & 16,662$^{***}$ &16,760$^{***}$ & \bf{16,607}
& 13.471$^{***}$ & 13.439$^{***}$ & 10.503$^{*}$ & 10.512$^{*}$ & \bf{10.487}
& 10.803$^{**}$ & 10.817$^{**}$ & 10.030$^{*}$ & 10.031$^{*}$ & \bf{10.025} \\

60 & 10.684$^{*}$ & 10.687$^{**}$ & 10.048$^{*}$ & 10.055$^{*}$ & \bf{10.037}
& 16,799$^{***}$ & 16,799$^{***}$ & 16,581$^{***}$ & 16,588$^{***}$ & \bf{16,567}
& 13.503$^{***}$ & 13.460$^{***}$ & 10.205$^{*}$  & 10.232$^{*}$ & \bf{10.105}
& 10.801$^{**}$ & 10.814$^{**}$ & 10.008$^{*}$ & 10.006$^{*}$ & \bf{10.005} \\

80 & 10.682$^{**}$ & 10.685$^{**}$ & 10.021$^{*}$ & 10.023$^{*}$ & \bf{10.015}
& 16,799$^{***}$ & 16,797$^{***}$ & 15,991$^{***}$ &16,552$^{***}$ & \bf{15,650}
& 13.469$^{***}$ & 13.555$^{***}$ & 10.067$^{*}$ & 10.097$^{*}$ & \bf{10.011}
& 10.800$^{**}$ & 10.804$^{***}$ & 10.001 & 10.002$^{*}$ & {10.001} \\

100 & 10.673$^{**}$ & 10.669$^{**}$ & 10.013$^{*}$ &10.016$^{*}$ & \bf{10.004}
& 16,796$^{***}$ & 16,797$^{***}$ & 13,628$^{***}$ & 16,486$^{***}$ & \bf{11,479}
& 13.518$^{***}$ & 13.453$^{***}$ & 10.012$^{*}$ &  10.040$^{*}$ & \bf{10}
& 10.797$^{**}$ & 10.790$^{**}$ & 10 &10 & 10 \\
\bottomrule

\multicolumn{13}{l}{\begin{minipage}{1.1\textwidth}  Notes. Best performers are bold. All experiments are repeated five times.

~~~~~~~~~~~~~~$^{*}p<0.1$; $^{**}p<0.05$; $^{***}p<0.001$.
\end{minipage}}
\end{tabular} }
\vspace{-5mm}
\end{table*}

\noindent \textbf{Comparison with Other Heuristics.}
\jiadong{As discussed in Section~\ref{sec:advanced-greedy}, the effectiveness (expected spread) of the AG algorithm is the same as the BG algorithm.
Thus, we compare the effectiveness (i.e., expected spread) of our proposed algorithms (AG and GR) with other heuristics (RA, OD and GS) on all networks (EC, F, W, EA, D, T, S, Y) under the IC model using pairwise t-tests in Tables~\ref{table:heuristics} (IMIN problem) and Table \ref{table:heuristics-eb} (IMIN-EB problem). The results under the LT model are included in the Tables A5 and A6 in online appendix. We first randomly select $10$ nodes as the seeds and vary the budget $b$ from $20,40,60,80$ to $100$. We repeat this process by $5$ times and report the average expected spread with the resulting blockers (the expected spread is computed by MCS with $10^5$ rounds). The bold numbers in Table \ref{table:heuristics} and \ref{table:heuristics-eb} denote the minimum expected spread across the four algorithms. The results under the WC model can be found in Table A4 and Table A5.}

The findings indicate that our GR algorithm consistently attains the optimal outcome, characterized by the lowest expected spread, across various budgets and propagation models. The performance of GR algorithm surpasses that of the other algorithms, with improvements that are statistically significant. Moreover, as the budget $b$ increases, the efficacy of the GR algorithm improves.
The findings substantiate the superiority of our proposed algorithm, GR, in terms of effectiveness when compared to other existing algorithms.

\begin{figure}[!htbp]
\begin{minipage}{0.49\linewidth}
\begin{subfigure}[h]{0.48\linewidth}
    \centering
    \includegraphics[width=\linewidth]{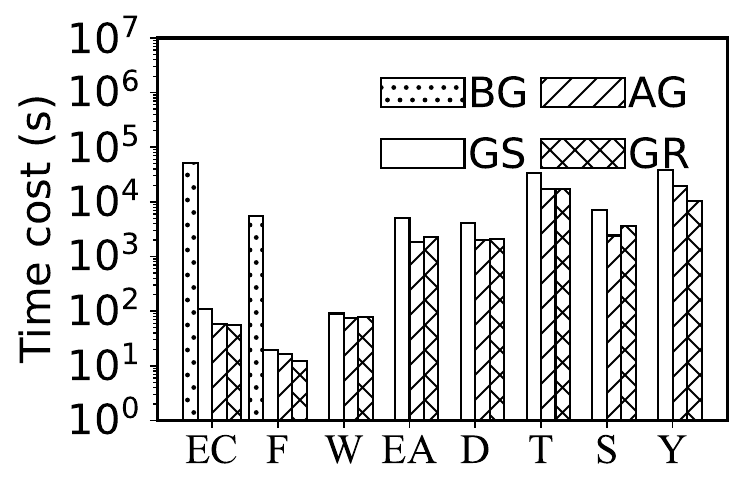}
    \caption{TR \& IC model in IMIN problem}
    \label{fig:time-tr}
    \end{subfigure}
    \begin{subfigure}[h]{0.48\linewidth}
    \centering
    \includegraphics[width=\linewidth]{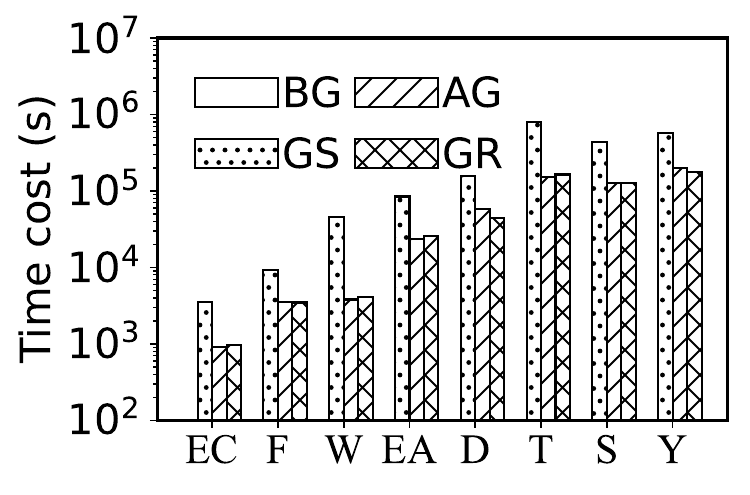}
    \caption{\jiadong{TR \& IC model in IMIN-EB problem}}
    \label{fig:time-tr-eb}
    \end{subfigure}
    \caption{Time Cost of Different Algorithms} \label{fig:time-imin}
    \end{minipage}
    \begin{minipage}{0.49\linewidth}
    \begin{subfigure}[h]{0.49\linewidth}
\centering
    \includegraphics[width=1\columnwidth]{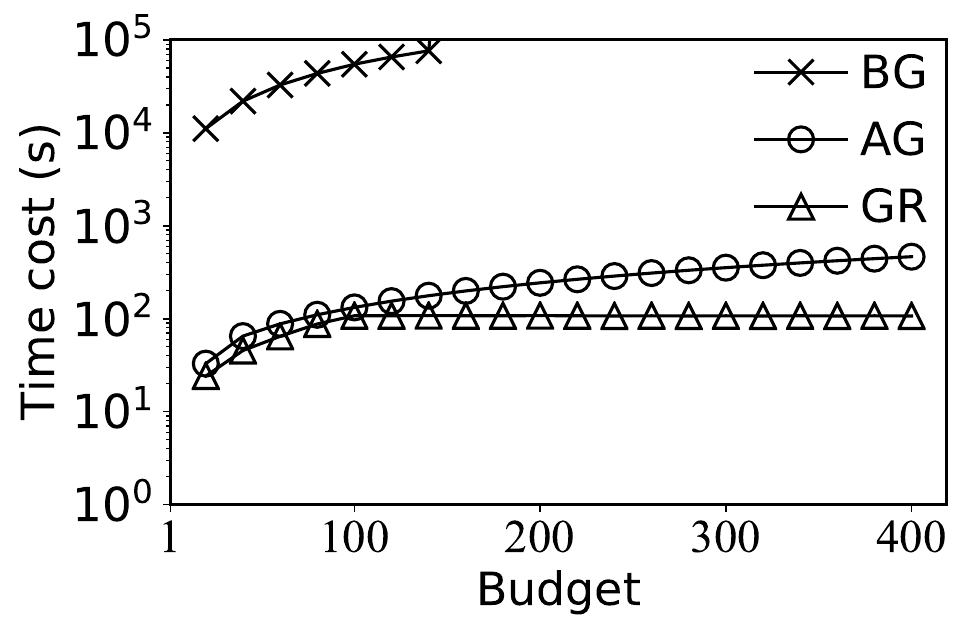}
    \caption{TR \& IC model in \texttt{Facebook} dataset}\label{fig:time-budget-a}
\end{subfigure}
\begin{subfigure}[h]{0.49\linewidth}
\centering
    \includegraphics[width=1\columnwidth]{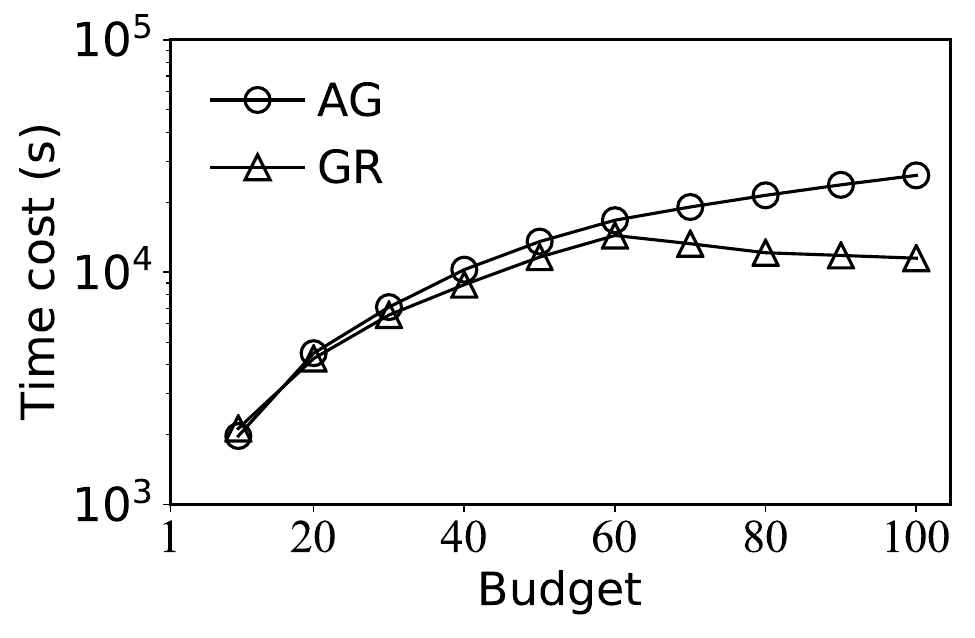}
    \caption{TR \& IC model in \texttt{DBLP} dataset}\label{fig:time-budget-c}
\end{subfigure}
\caption{Running Time v.s. Budget in the IMIN problem}
\label{fig:time-budget}
    \end{minipage}
    \vspace{-3mm}
\end{figure}

\noindent \textbf{Time Cost of Different Algorithms.}
\jiadong{We compare the running time of BG, GS, AG, and GR, and set the budget $b$ to $10$ due to the huge computation cost of the BG.
Figures~\ref{fig:time-imin}, A2 and A3 show the results for all networks under the two propagation models where the x-axis denotes different networks and the y-axis is the running times (s) of the algorithms.
For the IMIN problem, the BG algorithm is unable to return results within the specified time constraint of 24 hours for six networks under the TR model, and for five networks under the WC model. For IMIN-EB problem, the BG algorithm does not return any results within 72 hours in all datasets, while the number of candidates of IMIN-EB is much larger than IMIN (the number of edges v.s. the number of nodes).
The results indicate that our proposed algorithms, namely AG and GR algorithms, exhibit significantly better runtime performance compared to existing state-of-the-art BG algorithm. The difference in runtime is at least three orders of magnitude, and this gap tends to be even larger when applied to larger networks. These results align with the analysis of time complexities presented in Section~\ref{sec:advanced-greedy}. Additionally, the time required for GR is similar to that of AG.
We can also observe that the running time of all three algorithms under the LT model is generally smaller than under the IC model. This is likely due to the fact that the sampled graphs in the LT model are smaller, as each node can have at most one in-neighbor.
Additionally, the speedup in running time for the AG and GR algorithms is greater compared to the BG, which can be attributed to the strict linear refinement of these two algorithms under the LT model.}

{\color{rehigh}\noindent {\bf Varying the Budget.}
The running time of the algorithms on \texttt{Facebook} and \texttt{DBLP} networks are shown in Figures~\ref{fig:time-budget} and A4 for various budgets.
The running time of AG may decrease with larger budgets due to the early termination (Lines 19-20 in Algorithm~\ref{algo:greedyreplace}).
Our proposed algorithms AG and GR exhibit significantly greater efficiency in comparison to the most advanced existing method BG, with the disparity between them further enlarged as the budget expands. Additionally, it is observed that the computational time required for AG is similar to that of GR. AG may exhibit faster performance than GR in cases where the budget is limited. However, as the budget increases, GR demonstrates superior performance in terms of running time.



%% file: MS_Version/Appendix_MS.tex
\section*{Online Appendix 1: Theorems and Proofs}
\setcounter{table}{0}
\setcounter{figure}{0}
\renewcommand{\thetable}{A\arabic{table}}
\renewcommand{\thefigure}{A\arabic{figure}}
\subsection*{Proof of IMIN problem hardness}

\begin{theorem}
The IMIN problem is NP-Hard under the IC/LT model.
\label{theo::NP-IM}
\end{theorem}

\begin{proof}{Proof}
We reduce the densest k-subgraph (DKS) problem~\citep{BhaskaraCCFV10}, which is NP-hard, to the IMIN problem. Given an undirected graph $G(V,E)$ with $|V|=n$ and $|E|=m$, and an positive integer $k$, the DKS problem is to find a subset $A\subseteq V$ with exactly $k$ nodes such that the number of edges induced by $A$ is maximized.

\begin{figure}[!htbp]
    \centering
    \includegraphics[width=0.6\linewidth]{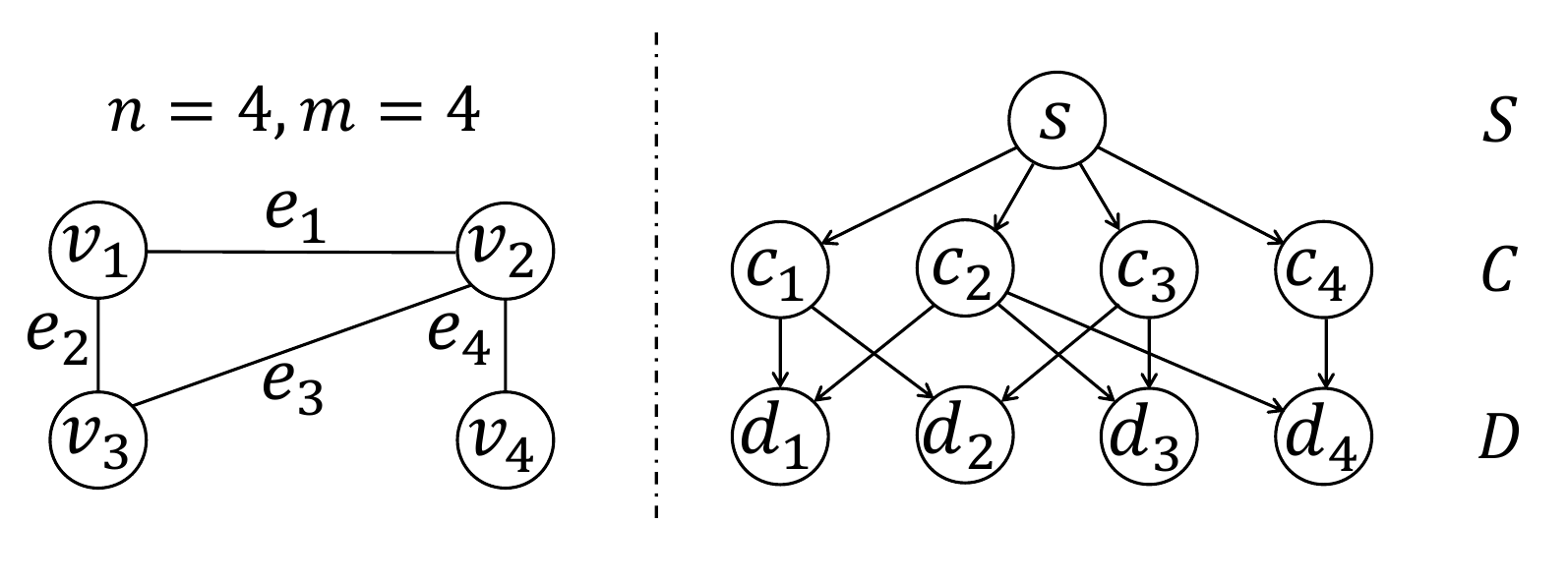}
    \caption{Construction example for hardness proofs, where the left part of the figure is an instance of DKS problem and the right part is its corresponding construction of IMIN problem.}
    \label{fig:np-hard}
\end{figure}

Consider an arbitrary instance $G(V=\{v_1,\cdots, v_n\},E=\{e_1,\cdots, e_{m}\})$ of DKS problem with a positive integer $k$, we construct a corresponding instance of IMIN problem on graph $G'$. Figure~\ref{fig:np-hard} shows a construction example from $4$ nodes and $4$ edges.

The graph $G'$ contains three parts: $C$, $D$, and a seed node $S$. The part $C$ contains $n$ nodes, i.e., $C=\cup_{1\le i\le n} c_i$ where each $c_i$ corresponds to $v_i$ of instance $G$. The part $D$ contains $m$ nodes, i.e., $D=\cup_{1\le i\le m} d_i$ where each node $d_i$ corresponds to edge $e_i$ in instance $G$. The node $S$ is the only seed of the graph.
For each edge $e_i=(v_x,v_y)$ in graph $G$, we add two edges: (i) from $c_x$ to $d_i$ and (ii) $c_y$ to $d_i$. Then we add an edge from $S$ to each $c_i (1\le i\le n)$. The propagation probability of each edge is set to $1$.
The construction of $G'$ is completed.

We then show the equivalence between the two instances.
As the propagation probability of each edge is $1$, the expected spread in the graph is equal to the number of nodes that can be reached by seed $S$.
Adding a node $v_i$ into node set $A$ corresponds to the removal of $c_i$ from the graph $G'$, i.e., $c_i$ is blocked. If edge $e_i$ is in the induced subgraph, the corresponding node $d_i$ in $G'$ cannot be reached by seed $S$. We find that blocking the nodes $c_i\in C$ will first lead to the decrease of expected spread of themselves, and the nodes in $D$ may also not be reached by $S$ if both two in-neighbors of them are blocked. Thus, blocking the corresponding nodes $c_i$ of $v_i\in A$ will lead to $|A|+g$ decrease of expected spread, where $g$ is the number of nodes in $D$ that cannot be reached by $S$ (equals to the number of edges in the induced subgraph $G[A]$). Note that there is no need to block nodes in $D$, because they do not have any out-neighbors, and blocking them only leads to the decrease of expected spread of themselves which is not larger than the decrease of the expected spread of blocking the nodes in $C$. 
{We can find that IMIN problem will always block $k$ nodes, as blocking one node will lead to at least $1$ decrease of expected spread.}

The optimal blocker set $B$ of IMIN problem corresponds to the optimal node set $A$ of DKS problem, where each node $c_i\in B$ corresponds to the set $v_i\in A$.
Thus, if the IMIN problem can be solved in PTIME, the DKS problem can also be solved in PTIME, while the DKS problem is NP-hard.\Halmos
\end{proof}

\subsection*{\jiadong{Proof of IMIN-EB problem hardness}}

\begin{theorem}
The IMIN-EB problem is NP-Hard under the IC/LT model.
\label{theo::imin-eb-np}
\end{theorem}

\begin{proof}{Proof}
\jiadong{We reduce the IMIN problem to IMIN-EB problem. Consider an arbitrary instance $G(V=\{v_1,\cdots, v_n\},E=\{e_1,\cdots, e_{m}\})$ of IMIN problem with a positive integer $b$, we construct a corresponding instance of IMIN-EB problem on graph $G'(V',E')$. There are $2n$ nodes in graph $G'$, i.e., $V'=\cup _{1\le i\le n} \{v_i^{in},v_i^{out}\}$ where each pair of nodes $\{v_i^{in},v_i^{out}\}$ corresponds to node $v_i$ in instance $G$. For each edge $e_i=(v_x,v_y)$ in graph $G$, we add edge from $v_x^{out}$ to $v_y^{in}$ in graph $G'$. We further add edges from $v_i^{in}$ to $v_i^{out}$ for each pair $\{v_i^{in},v_i^{out}\}$ with probability $1$.
We can find that the decrease of the expected spread led by $(v_x^{out},v_y^{in})$ is always less than or equal to $(v_x^{in},v_x^{out})$, thus we always block edges from $v_x^{in}$ to $v_x^{out}$, which is equal to block node $v_x$ in the IMIN problem. Thus the IMIN-EB problem can not be solved in PTIME.}
\Halmos \end{proof}

\subsection*{Proof of Corollary~\ref{corollary::es-onenode}}

\begin{proof}{Proof}
\revise{Let $g'=g[V(g)\setminus \{u\}]$, we have $\sigma(s,g)-\sigma^{\rightarrow u}(s,g)=\sigma(s,g')$. Thus, $\mathbb{E}[\sigma(s,g)-\sigma^{\rightarrow u}(s,g)]=\mathbb{E}[\sigma(s,g')]$. As $g'=g[V(g)\setminus \{u\}]$, $g'$ can be regarded as a random sampled graph derived from graph $G[V\setminus \{u\}]$. By Lemma~\ref{lemma:sample}, $\mathbb{E}[\sigma(s,g')]=\mathbb{E}(\{s\},G[V\setminus \{u\}])$. Therefore, $\mathbb{E}[\sigma(s,g)-\sigma^{\rightarrow u}(s,g)]=\mathbb{E}(\{s\},G[V\setminus \{u\}])$.}
\Halmos \end{proof}

\subsection*{Proof of Theorem~\ref{theo::block_expected}}

\begin{proof}{Proof}
\sloppy
\revise{
Since $\sigma^{\rightarrow u}(s,g)=\sigma(s,g)-(\sigma(s,g)-\sigma^{\rightarrow u}(s,g))$, we have $\mathbb{E}[\sigma^{\rightarrow u}(s,g)]=\mathbb{E}[\sigma(s,g)]-\mathbb{E}[\sigma(s,g)-\sigma^{\rightarrow u}(s,g)]$. 
Then from Corollary~\ref{corollary::es-onenode}, we have $\mathbb{E}[\sigma(s,g)-\sigma^{\rightarrow u}(s,g)]=\mathbb{E}(\{s\},G[V\setminus \{u\}])$.
By Lemma~\ref{lemma:sample}, we know $\mathbb{E}[\sigma(s,g)]=\mathbb{E}(\{s\},G)$.
Therefore, combine them we know $\mathbb{E}[\sigma^{\rightarrow u}(s,g)]=\mathbb{E}(\{s\},G)-\mathbb{E}(\{s\},G[V\setminus \{u\}])$.}
\Halmos \end{proof}

{The lemma of Chernoff bounds~\citep{MotwaniR95}.

\begin{lemma}
Let $X$ be the sum of $c$ i.i.d. random variables sampled from a distribution on $[0,1]$ with a mean $\mu$. For any $\delta >0$, we have $Pr[X-c\mu \ge \delta \cdot c\mu] \le e^{-\delta^2c\mu /(2+\delta)}$ and $Pr[X-c\mu \le -\delta \cdot c\mu] \le e^{-\delta^2c\mu /2}.$
\label{lemma:cher}
\end{lemma}}

\subsection*{Proof of Theorem~\ref{theorem:approx}}

\begin{proof}{Proof}
We have
$Pr[|\xi^{\rightarrow u}(s,G)-OPT|\ge \varepsilon\cdot OPT]$
$
\begin{matrix}
 =& Pr[|\frac{\xi^{\rightarrow u}(s,G)\cdot \theta}{n}-\frac{OPT\cdot \theta}{n}|\ge \frac{\varepsilon \cdot \theta}{n}\cdot OPT] \\
\end{matrix}
$.

Let $\delta=\varepsilon$, $\mu = \frac{OPT}{n}$ and $X=\frac{\xi^{\rightarrow u}(s,G)\cdot \theta}{n}$.
According to Lemma~\ref{lemma:cher}, we have
$Pr[|\xi^{\rightarrow u}(s,G)-OPT|\ge \varepsilon\cdot OPT]$
$=  Pr[|X-\theta \mu|\ge \delta \cdot \theta \mu] 
\le exp(\frac{-\delta^2\theta\mu}{2+\delta})
 = exp(\frac{-\varepsilon^2\theta OPT}{n(2+\varepsilon)})$.


As $\theta\ge \frac{l(2+\varepsilon)n\log n}{\varepsilon^2\cdot OPT}$, we have $Pr[|\xi^{\rightarrow u}(s,G)-OPT|\ge \varepsilon\cdot OPT]\le exp(l \log n) = n^{-l}$.
Therefore, $|\xi^{\rightarrow u}(s,G)-OPT| < \varepsilon\cdot OPT$ holds with at least $1-n^{-l}$ probability.
\Halmos \end{proof}

\subsection*{Proof of Theorem~\ref{theo:dominator-subtree}}

\begin{proof}{Proof}
\revise{Let $c[u]$ denote the size of the subtree with root $u$ in the dominator tree of graph $g$. Assume $v$ is in the subtree with $u$ (i.e., $u$ is the ancestor of $v$), from the definition of dominator, we have $v$ cannot be reached by $s$ when $u$ is blocked. Thus, $c[u]\le \sigma^{\rightarrow u}(s,g)$. If $v$ is not in the subtree, i.e., $u$ does not dominate $v$ in the graph, there is a path from $s$ to $v$ not through $u$, which means blocking $u$ will not affect the reachability from $s$ to $v$. We have $\sigma^{\rightarrow u}(s,g) =  c[u]$.}
\Halmos \end{proof}

\subsection*{Proof of Theorem~\ref{theo::edge_expected}}

\begin{proof}{Proof}
As only nodes in $G$ have activation probability, based on Lemma~\ref{lemma:sample}, the expected number of nodes in $V_V$ is equal to the expected spread of $G$.
We can find that blocking the edge $(u,v)$ in graph $G$ will remove the node $w_{u,v}$ from graph $g^E$.
From Theorem~\ref{theo::block_expected}, we have the decrease of expected spread by blocking edge $(u,v)$ in $G$ is equal to the decrease of expected spread by blocking node $w_{u,v}$ in $g^E$, i.e., the expected number of nodes in $R_{u,v}$.
\Halmos \end{proof}

\subsection*{Proof of Theorem~\ref{theo:edge-dominator-subtree}}

\begin{proof}{Proof}
From Theorem~\ref{theo:dominator-subtree}, we know nodes satisfying (a) reachable from $s$ in $g^E$ and (b) all the paths from $s$ in $g^E$ pass through $w_{u,v}$ constitute the subtree rooted at $w_{u,v}$. $R_{u,v}$ further requires nodes are in the $V_V$, and thus $R_{u,v}$ contains the nodes that both in $V_V$ and subtree rooted at $w_{u,v}$. 
\Halmos \end{proof}

\subsection*{Proof of the function of expected spread under the IC model in IMIN problem}

{\color{rehigh}\begin{theorem}
Given a graph $G$ and a seed set $S$, the expected spread function $\mathbb{E}(S,G[V \setminus B])$ is monotone and not supermodular of $B$ under the IC model.
\label{theo::expectedspread}
\end{theorem}

\begin{proof}{Proof}
As adding any blocker to any set $B$ cannot increase the expected influence spread, we have $\mathbb{E}(S,G[V \setminus B])$ is monotone of $B$.
For every two set $X\subseteq Y\subseteq V$ and node $x\in V\setminus Y$, if function $f(\cdot)$ is supermodular, it must hold that $f(X\cup \{x\})-f(X)\le f(Y\cup \{x\})-f(Y)$. 
Consider the graph in Figure~\ref{fig:sample}, let $f(B)=\mathbb{E}(S,G[V \setminus B])$, $X=\{v_3\},Y=\{v_2,v_3\}$ and $x=v_4$. As $f(X)=6.66$, $f(Y)=5.66$, $f(X\cup \{x\})=5.66$, and $f(Y\cup \{x\})=1$, we have $f(X\cup \{x\})-f(X)=-1 > f(Y\cup \{x\})-f(Y)=-4.66$.
\Halmos \end{proof}



\subsection*{Proof of Approximation Hardness under the IC Model for IMIN problem}

\begin{theorem}
Under the IC model, the IMIN problem is APX-hard unless P=NP when $\frac{n+m-b}{OPT}=c$, where $OPT$ is the optimal result of the IMIN problem.
\label{theo::APX}
\end{theorem}

\begin{proof}{Proof}
We use the same reduction from the densest k-subgraph (DKS) problem to the Influence Minimization problem, as in the proof of Theorem~6.
Densest k-subgraph (DKS) problem does not have any polynomial-time approximation scheme, unless P=NP~\citep{Khot06}.
According to the proof of Theorem~6, we have a blocker set $B$ for influence minimization problem on $G'$ corresponding to a node set $A$ for DKS problem, where each $c_i\in B$ corresponds to $v_i\in A$.
Let $x$ denote the number of edges in the optimal result of DKS, and $y$ denote the optimal spread in IMIN, we have $x+k+y=1+n+m$, where $k$ is the given positive number of DKS problem.
If there is a solution with $\gamma$-approximation on the influence minimization problem, there will be a $\lambda=\frac{c-\lambda}{c-1}$-approximation on the DKS problem.
\Halmos \end{proof}


\subsection*{Proof of the function of expected spread under the LT model in IMIN problem}


\jiadong{
\begin{theorem}
Given a graph $G$ and a seed set $S$, the expected spread function $\mathbb{E}(S,G[V \setminus B])$ is monotone and supermodular of $B$ under the LT model.
\label{theo::expectedspread-LT}
\end{theorem}}


\jiadong{
As the expected spread is non-increasing and supermodular of blocker set $B$ (Theorem~10), the AdvancedGreedy algorithm gives a $(1-1/e-\varepsilon)$-approximation to the optimum for IMIN problem under the LT model~\citep{tzoumas2016sensor}.}

\begin{proof}{Proof}
\jiadong{
As adding any blocker to any set $B$ cannot increase the expected influence spread, we have $\mathbb{E}(S,G[V \setminus B])$ is monotone of $B$.
In the sampled graphs under the LT model, we know that the immediate dominator of each node is its unit in-neighbor, thus the children in the dominator tree of each node are the nodes that are reachable from it. Therefore, blocking any node $u$ in the graph leads to remove the edge $(fa_u,u)$ in the dominator tree, where $fa_u$ denotes the father node of node $u$ in the dominator tree. We can find that, for any node $u$ in each sampled graph, adding any blocker to any set $B$ cannot increase the number of nodes $u$.
Let $f(B)=\mathbb{E}(S,G[V \setminus B])$, we have for any two set $X,Y(X\subseteq Y\subseteq V)$ and a node $u\in V$, $f(X)-f(X\cup\{u\})\ge f(Y)-f(Y\cup\{u\})$.}
\Halmos \end{proof}



\subsection*{Theorems of function of expected spread and the approximation hardness under the IC/LT model in IMIN-EB problem}


\jiadong{
\begin{theorem}
Given a graph $G$ and a seed set $S$, the expected spread function $\mathbb{E}(S,G(V, E \setminus B))$ is monotone and not supermodular of $B$ under the IC model.
\label{theo::expectedspread-edge}
\end{theorem}}

\jiadong{
\begin{theorem}
Given a graph $G$ and a seed set $S$, the expected spread function $\mathbb{E}(S,G(V, E \setminus B))$ is monotone and supermodular of $B$ under the LT model.
\label{theo::expectedspread-edge-LT}
\end{theorem}}

\jiadong{
Therefore, as the expected spread is non-increasing and supermodular of blocked edge set $B$ (Theorem~12), the AdvancedGreedy algorithm gives a $(1-1/e-\varepsilon)$-approximation to the optimum for IMIN-EB problem under the LT model~\citep{tzoumas2016sensor}.}



\newpage
\section*{Online Appendix 2: Psuedo-codes}
\setcounter{algocf}{0}
\renewcommand{\thealgocf}{A\arabic{algocf}}
\begin{algorithm}[htp]
    \SetVline 
    \SetFuncSty{textsf}
    \SetArgSty{textsf}
	\caption{BaselineGreedy($G,s$)}
	\label{algo:greedy}
	\Input{a graph $G$ and the source $s$}
	\Output{the blocker (resp. edge) set $B$}
	\State{$B\leftarrow $ empty}
	\For{$i\leftarrow 1$ to $b$}
	{
	    \State{$x\leftarrow -1$}
	    \For{each $u\in V(G)\setminus (B\cup \{s\})$ (resp. each $(u,v)\in E(G)\setminus B$)}
	    {
	        \State{$\Delta \leftarrow$ decrease of expected spread when blocking $u$ (resp. $(u,v)$)}
	        \If{$x=-1$ or $\Delta > \Delta[x]$}
	        {
	            \State{$x\leftarrow u$ (resp. $x\leftarrow (u,v)$)}
	        }
	    }
	    \State{$B\leftarrow B\cup \{x\}$}
	}
	\Return{$B$}
\end{algorithm}

\begin{algorithm}[htp]
    \SetVline 
    \SetFuncSty{textsf}
    \SetArgSty{textsf}
	\caption{\jiadong{DTC-LT($G,s$)}}
	\label{algo:tree-LT}
	\Input{\jiadong{a sampled graph under the LT model $G$, and the source $s$}}
	\Output{\jiadong{$DT$ dominator tree of graph $G$}}
        \State{\jiadong{Push $s$ to the end of the queue $Q$}}
        \State{\jiadong{$vis\leftarrow \{s\},DT\leftarrow \varnothing$}}
        \While{\jiadong{$Q$ is not empty}}
        {
            \State{\jiadong{$v\leftarrow$ $Q$.pop()}}
            \For{\jiadong{each $u\in N^{out}_v$}}
            {
                \If{\jiadong{$u \not\in vis$}}
                {
                    \State{\jiadong{$vis\leftarrow vis\cup \{u\}$}}
                    \State{\jiadong{Push $u$ to the end of the queue $Q$}}
                    \State{\jiadong{add edge $(v,u)$ into $DT$}}
                }
            }
        }
        \State{\jiadong{\Return{$DT$}}}
\end{algorithm}

\newpage
\section*{Online Appendix 3: Tables}

\begin{table}[htp]
    \caption{Summary of Notations}
    \small
    \label{tab:nota}
       \centering
       \resizebox{\linewidth}{!}{
       \begin{tabular}{|p{0.18\columnwidth}|p{0.74\columnwidth}|}\hline
        Notation & Definition\\ \hline\hline
       $G=(V,E)$ & a directed graph with node set $V$ and edge set $E$\\ \hline
       $n; m$ & number of nodes/edges in $G$ (assume $m > n$)\\ \hline
       $V(G); E(G)$ & the set of nodes/edges in $G$\\ \hline
       $G[V']$ & the subgraph in $G$ induced by node set $V'$\\ \hline
       $N^{in}_u; N^{out}_u$ & the set of in-neighbors/out-neighbors of node $u$ \\ \hline
       $d^{in}_u; d^{out}_u$ & the in-degree/out-degree of node $u$ \\ \hline
       $S; s$ & the seed set; a seed node \\ \hline
       $B$ & the blocker set\\ \hline
       $\theta$ & the number of sampled graphs used in algorithm \\ \hline
       $Pr[x]$ & the probability if $x$ is true\\ \hline
       $\mathbb{E}[x]$ & the expectation of variable $x$\\ \hline
       $p_{u,v}$ & the probability that node $u$ activates node $v$\\ \hline
       $\mathcal{P}^G(x,S)$ & the probability that node $x$ is activated by set $S$ in $G$\\ \hline
       $\mathbb{E}(S,G)$ & the expected influence spread, i.e., the expected number of activated non-seed nodes in graph $G$ from seed set $S$ \\ \hline
       \end{tabular}}
\end{table}



\begin{table}[htp]
    \centering
\caption{\jiadong{Exact v.s. GreedyReplace (IC model)}}\label{tab:exact-wc}
\resizebox{0.8\linewidth}{!}{\color{jiadong}
\begin{tabular}{|c|c|c|c|c|c|c|c|c|c|c|}
\toprule
\multirow{2}{*}{ b } & \multicolumn{3}{c|}{Expected Spread} & \multicolumn{2}{c|}{Running Time (s)} & \multicolumn{3}{c|}{Expected Spread} & \multicolumn{2}{c|}{Running Time (s)}  \\ \cline{2-11}
& Exact & GR & Ratio & Exact & GR & Exact & GR & Ratio & Exact & GR \\ 
\midrule
& \multicolumn{5}{c|}{Panel C: IMIN Problem (WC Model)} & \multicolumn{5}{c|}{Panel D: IMIN-EB Problem (WC Model)}\\
\midrule
1 &  11.185 & 11.185 & 100\% & 2.63 & 0.10 & 11.882 & 11.882 & 100\% & 4.03 & 0.12 \\
2 &  11.077 & 11.078 & 99.99\% & 110.92 & 0.18 & 11.634 & 11.697 & 99.5\% & 205.21 & 0.25  \\
3 &  10.997 & 10.998 & 99.99\% & 3284.0 & 0.23 & 11.344 & 11.421 & 99.3\% & 7375.2 & 0.37  \\
4 &  10.922 & 10.925 & 99.97\% & 69415 & 0.33 & 11.269 & 11.279 & 99.9\% & 98247 & 0.50 \\
\midrule
\bottomrule
\end{tabular}
}
\end{table}


\begin{table*}[htp]
\centering
\caption{\jiadong{Comparision with Other Heuristics (Expected Spread) in IMIN problem (IC model)}}\label{table:heuristics-wc}
\resizebox{0.9\linewidth}{!}{
\begin{tabular}{c|llll|llll|llll|llll}

\toprule
\multirow{2}{*}{\centering b } & \multicolumn{4}{c|}{EmailCore (WC model)} & \multicolumn{4}{c|}{Facebook (WC model)} & \multicolumn{4}{c|}{Wiki-Vote (WC model)} & \multicolumn{4}{c}{EmailAll (WC model)} \\
& RA & OD & AG & GR & RA & OD & AG & GR & RA & OD & AG & GR & RA & OD & AG & GR \\
\midrule
20 & 82.605$^{***}$ & 54.907$^{**}$ & 53.516$^{*}$ & \bf{53.296} & 21.482$^{***}$ & 21.362$^{***}$ & 14.588$^{*}$ & \bf{14.554} & 24.102$^{***}$ & 22.660$^{***}$ & 17.765$^{*}$ & \bf{17.701} & 13.330$^{***}$ & 11.493$^{**}$ & 10.720$^{*}$ & \bf{10.455} \\
40 & 75.990$^{***}$ & 44.710$^{***}$ & 40.199$^{*}$ & \bf{40.093} & 21.456$^{***}$ & 21.360$^{***}$ & 12.425$^{*}$ & \bf{12.418} & 23.971$^{***}$ & 21.696$^{***}$ & 15.258$^{*}$ & \bf{15.222} & 13.234$^{***}$ & 11.447$^{**}$ & 10.111$^{*}$ & \bf{10.086} \\
60 & 69.947$^{***}$ & 37.561$^{***}$ & 31.891$^{*}$ & \bf{31.784} & 21.429$^{***}$ & 21.297$^{***}$ & 11.194$^{*}$ & \bf{11.187} & 23.899$^{***}$ & 20.409$^{***}$ & 13.749$^{*}$ & \bf{13.743} & 13.217$^{***}$ & 11.408$^{***}$ & 10 & 10 \\
80 & 64.154$^{***}$ & 32.580$^{***}$ & 26.094$^{*}$ & \bf{26.073} & 21.417$^{***}$ & 21.176$^{***}$ & 10.476$^{*}$ & \bf{10.474} & 23.763$^{***}$ & 12.798$^{*}$ & 12.654$^{*}$ & \bf{12.574} & 13.188$^{***}$ & 11.385$^{**}$ & 10 & 10 \\
100 & 57.170$^{***}$ & 24.959$^{***}$ & 21.926$^{*}$ & \bf{21.899} & 21.395$^{***}$ & 21.056$^{***}$ & 10.013$^{*}$ & \bf{10.012} & 23.757$^{***}$ & 12.711$^{*}$ & 12.138$^{*}$ & \bf{12.129} & 13.070$^{***}$ & 11.324$^{**}$ & 10 & 10 \\
\bottomrule

\toprule
\multirow{2}{*}{\centering b } & \multicolumn{4}{c|}{DBLP (WC model)} & \multicolumn{4}{c|}{Twitter (WC model)} & \multicolumn{4}{c|}{Stanford (WC model)} & \multicolumn{4}{c}{Youtube (WC model)} \\ 
& RA & OD & AG & GR & RA & OD & AG & GR & RA & OD & AG & GR & RA & OD & AG & GR \\
\midrule
20 & 118.23$^{***}$ & 118.17$^{***}$ & 32.602$^{*}$ & \bf{32.601} & 259.45$^{***}$ & 235.73$^{***}$ & 199.40$^{*}$ & \bf{198.76} & 25.803$^{***}$ & 25.800$^{***}$ & 11.742$^{*}$ & \bf{11.740} & 25.663$^{***}$ & 25.368$^{***}$ & 10.152$^{*}$ & \bf{10.113} \\
40 & 118.16$^{***}$ & 117.81$^{***}$ & 18.429$^{*}$ & \bf{18.409} & 258.86$^{***}$ & 226.28$^{***}$ & 170.20$^{**}$ & \bf{168.64} & 25.790$^{***}$ & 25.779$^{***}$ & 10.435$^{*}$ & \bf{10.398} & 25.585$^{***}$ & 25.330$^{***}$ & 10.012$^{*}$ & \bf{10.006} \\
60 & 118.09$^{***}$ & 117.73$^{***}$ & 11.869$^{*}$ & \bf{11.867} & 257.75$^{***}$ & 214.90$^{***}$ & 144.61$^{*}$ & \bf{144.23} & 25.777$^{***}$ & 25.771$^{***}$ & 10.119$^{*}$ & \bf{10.101} & 25.446$^{***}$ & 25.149$^{***}$ & 10 & 10 \\
80 & 117.97$^{***}$ & 117.59$^{***}$ & 10 & 10 & 255.48$^{***}$ & 204.17$^{***}$ & 129.53$^{*}$ & \bf{128.75} & 25.685$^{***}$ & 25.633$^{***}$ & 10.005$^{*}$ & \bf{10.002} & 25.346$^{***}$ & 25.113$^{***}$ & 10 & 10 \\
100 & 117.94$^{***}$ & 117.43$^{***}$ & 10 & 10 & 254.04$^{***}$ & 196.40$^{***}$ & 114.11$^{**}$ & \bf{112.07} & 25.657$^{***}$ & 25.603$^{***}$ & 10.002$^{*}$ & \bf{10.000} & 25.277$^{***}$ & 25.052$^{***}$ & 10 & 10 \\
\bottomrule
\multicolumn{13}{l}{\begin{minipage}{1.1\textwidth}  Notes. Best performers are bold. All experiments are repeated five times.

~~~~~~~~~~~~~~$^{*}p<0.1$; $^{**}p<0.05$; $^{***}p<0.001$.
\end{minipage}} 
\end{tabular}}
\end{table*}

\begin{table*}[htp]
\centering
\caption{\jiadong{Comparision with Other Heuristics (Expected Spread) in IMIN-EB problem (IC model)}}\label{table:heuristics-eb-wc}
\resizebox{0.9\linewidth}{!}{
\begin{tabular}{c|llll|llll|llll|llll}

\toprule
\multirow{2}{*}{\centering b } & \multicolumn{4}{c|}{EmailCore (WC model)} & \multicolumn{4}{c|}{Facebook (WC model)} & \multicolumn{4}{c|}{Wiki-Vote (WC model)} & \multicolumn{4}{c}{EmailAll (WC model)} \\
& RA & OD & AG & GR & RA & OD & AG & GR & RA & OD & AG & GR & RA & OD & AG & GR \\
\midrule
20 & 110.48$^{***}$ & 109.56$^{***}$ & 98.538$^{**}$ & \bf{96.303} & 34.122$^{***}$ & 34.679$^{***}$ & 18.245$^{*}$ & \bf{17.784} & 14.559$^{***}$ & 14.549$^{***}$ & 12.633$^{*}$ & \bf{12.585} & 135.77$^{***}$ & 134.68$^{***}$ & 112.28$^{***}$ & \bf{96.303} \\
40 & 108.94$^{***}$ & 109.28$^{***}$ & 90.809$^{*}$ & \bf{90.303} & 34.084$^{***}$ & 34.415$^{***}$ & 15.200$^{*}$ & \bf{15.146} & 14.517$^{***}$ & 14.449$^{***}$ & 11.685$^{*}$ & \bf{11.668} & 135.52$^{***}$ & 134.60$^{***}$ & 93.690$^{**}$ & \bf{90.327} \\
60 & 107.77$^{***}$ & 108.26$^{***}$ & 83.598$^{*}$ & \bf{82.780} & 34.006$^{***}$ & 33.979$^{***}$ & 13.721$^{*}$ & \bf{13.638} & 14.462$^{***}$ & 14.448$^{***}$ & 11.063$^{*}$ & \bf{11.041} & 135.34$^{***}$ & 134.46$^{***}$ & 83.598$^{***}$ & \bf{75.873} \\
80 & 106.78$^{***}$ & 107.77$^{***}$ & 78.104$^{*}$ & \bf{77.089} & 33.779$^{***}$ & 33.974$^{***}$ & 12.842$^{*}$ & \bf{12.811} & 14.449$^{***}$ & 14.416$^{***}$ & 10.618$^{*}$ & \bf{10.603} & 134.68$^{***}$ & 133.74$^{***}$ & 55.483$^{*}$ & \bf{54.509} \\
100 & 106.49$^{***}$ & 106.25$^{***}$ & 73.944$^{**}$ & \bf{71.778} & 33.775$^{***}$ & 33.952$^{***}$ & 12.160$^{*}$ & \bf{12.127} & 14.446$^{***}$ & 14.344$^{***}$ & 10.303$^{*}$ & \bf{10.285} & 133.65$^{***}$ & 133.18$^{***}$ & 37.619$^{*}$ & \bf{37.383} \\
\bottomrule

\toprule
\multirow{2}{*}{\centering b } & \multicolumn{4}{c|}{DBLP (WC model)} & \multicolumn{4}{c|}{Twitter (WC model)} & \multicolumn{4}{c|}{Stanford (WC model)} & \multicolumn{4}{c}{Youtube (WC model)} \\ 
& RA & OD & AG & GR & RA & OD & AG & GR & RA & OD & AG & GR & RA & OD & AG & GR \\
\midrule
20 & 12.943$^{**}$ & 12.957$^{**}$ & 11.915$^{*}$ & \bf{11.531} & 182.89$^{***}$ & 181.91$^{***}$ & 165.14$^{*}$ & \bf{164.771} & 30.707$^{***}$ & 30.503$^{***}$ & 16.758$^{*}$ & \bf{16.717} & 36.406$^{***}$ & 36.428$^{***}$ & 26.471$^{**}$ & \bf{25.339} \\
40 & 12.932$^{***}$ & 12.948$^{***}$ & 10.963$^{*}$ & \bf{10.834} & 181.96$^{***}$ & 180.82$^{***}$ & 157.06$^{**}$ & \bf{155.303} & 30.572$^{***}$ & 30.495$^{***}$ & 14.427$^{*}$ & \bf{14.120} & 36.393$^{***}$ & 36.383$^{***}$ & 21.303$^{*}$ & \bf{20.353} \\
60 & 12.928$^{***}$ & 12.929$^{***}$ & 10.725$^{*}$ & \bf{10.596} & 181.43$^{***}$ & 180.02$^{***}$ & 150.74$^{*}$ & \bf{150.266} & 30.564$^{***}$ & 30.225$^{***}$ & 12.495$^{*}$ & \bf{12.489} & 36.373$^{***}$ & 36.368$^{***}$ & 16.358$^{*}$ & \bf{15.433} \\
80 & 12.906$^{***}$ & 12.927$^{***}$ & 10.398$^{*}$ & \bf{10.319} & 180.44$^{***}$ & 179.73$^{***}$ & 145.82$^{*}$ & \bf{144.653} & 30.292$^{***}$ & 29.894$^{***}$ & 10.971$^{*}$ & \bf{10.262} & 36.370$^{***}$ & 36.358$^{***}$ & 11.407$^{**}$ & \bf{10.385} \\
100 & 12.895$^{***}$ & 12.922$^{***}$ & 10.146$^{*}$ & \bf{10.070} & 180.08$^{***}$ & 178.95$^{***}$ & 141.99$^{*}$ & \bf{141.681} & 30.034$^{***}$ & 29.693$^{***}$ & 10.187$^{*}$ & \bf{10.136} & 36.363$^{***}$ & 36.319$^{***}$ & 10 & 10 \\
\bottomrule
\multicolumn{13}{l}{\begin{minipage}{1.1\textwidth}
Notes. Best performers are bold. All experiments are repeated five times.

~~~~~~~~~~~~~~$^{*}p<0.1$; $^{**}p<0.05$; $^{***}p<0.001$.\end{minipage}}
\end{tabular} }
\end{table*}

\begin{table*}[htp]
\centering
\caption{\jiadong{Comparision with Other Heuristics (Expected Spread) in IMIN problem (LT model)}}\label{table:heuristics-eb-wc-lt}
\resizebox{0.9\linewidth}{!}{
\begin{tabular}{c|lll|lll|lll|lll}
\toprule
\multirow{2}{*}{\centering b } & \multicolumn{3}{c|}{EmailCore (WC model)} & \multicolumn{3}{c|}{Facebook (WC model)} & \multicolumn{3}{c|}{Wiki-Vote (WC model)} & \multicolumn{3}{c}{EmailAll (WC model)} \\
& RA & OD & AG  & RA & OD & AG  & RA & OD & AG  & RA & OD & AG  \\ \midrule
20 & 111.63$^{***}$ & 74.122$^{**}$ & \bf{72.291} & 29.741$^{***}$ & 29.289$^{***}$ & \bf{17.841} & 30.751$^{***}$ & 28.610$^{***}$ & \bf{20.87} & 16.109$^{***}$ & 13.015$^{***}$ & \bf{11.743} \\
40 & 94.271$^{***}$ & 61.501$^{***}$ & \bf{49.055} & 28.911$^{***}$ & 28.978$^{***}$ & \bf{14.041} & 30.076$^{***}$ & 27.910$^{***}$ & \bf{18.658} & 15.867$^{***}$ & 12.738$^{***}$ & \bf{10.66} \\
60 & 93.424$^{***}$ & 47.736$^{***}$ & \bf{41.895} & 28.492$^{***}$ & 26.821$^{***}$ & \bf{11.869} & 29.139$^{***}$ & 25.914$^{***}$ & \bf{15.749} & 15.791$^{***}$ & 12.147$^{***}$ & \bf{10} \\
80 & 84.402$^{***}$ & 41.032$^{***}$ & \bf{31.894} & 28.116$^{***}$ & 26.468$^{***}$ & \bf{11.85} & 28.407$^{***}$ & 15.064$^{***}$ & \bf{11.113} & 15.689$^{***}$ & 12.111$^{**}$ & \bf{10} \\
100 & 77.613$^{***}$ & 33.140$^{***}$ & \bf{30.217} & 26.818$^{***}$ & 26.061$^{***}$ & \bf{10.423} & 27.474$^{***}$ & 13.757$^{***}$ & \bf{10.744} & 15.349$^{***}$ & 12.066$^{**}$ & \bf{10} \\
\bottomrule
\toprule
\multirow{2}{*}{\centering b } & \multicolumn{3}{c|}{DBLP (WC model)} & \multicolumn{3}{c|}{Twitter (WC model)} & \multicolumn{3}{c|}{Stanford (WC model)} & \multicolumn{3}{c}{Youtube (WC model)} \\ 
& RA & OD & AG  & RA & OD & AG  & RA & OD & AG  & RA & OD & AG  \\  \midrule
20 & 164.57$^{***}$ & 164.65$^{***}$ & \bf{39.367} & 350.20$^{***}$ & 327.51$^{***}$ & \bf{265.73} & 28.027$^{***}$ & 27.809$^{***}$ & \bf{12.866} & 28.121$^{***}$ & 27.375$^{***}$ & \bf{10.835} \\
40 & 162.24$^{***}$ & 159.38$^{***}$ & \bf{23.893} & 348.53$^{***}$ & 308.57$^{***}$ & \bf{212.88} & 27.971$^{***}$ & 27.379$^{***}$ & \bf{11.301} & 27.710$^{***}$ & 26.572$^{***}$ & \bf{10.058} \\
60 & 161.07$^{***}$ & 148.66$^{***}$ & \bf{12.517} & 347.07$^{***}$ & 271.03$^{***}$ & \bf{194.46} & 26.882$^{***}$ & 26.833$^{***}$ & \bf{10.680} & 26.921$^{***}$ & 26.057$^{***}$ & \bf{10} \\
80 & 145.60$^{***}$ & 147.03$^{***}$ & \bf{10} & 335.84$^{***}$ & 262.19$^{***}$ & \bf{166.57} & 26.614$^{***}$ & 26.319$^{***}$ & \bf{10.101} & 26.190$^{***}$ & 25.956$^{***}$ & \bf{10} \\
100 & 143.88$^{***}$ & 144.71$^{***}$ & \bf{10} & 321.43$^{***}$ & 254.65$^{***}$ & \bf{158.14} & 26.386$^{***}$ & 25.983$^{***}$ & \bf{10.076} & 26.101$^{***}$ & 25.409$^{***}$ & \bf{10} \\
\bottomrule
\multicolumn{13}{l}{\begin{minipage}{1.1\textwidth}  Notes. Since the LT model requires the probabilities of incoming edges for each node to be equal to or less than $1$, we choose to evaluate our algorithms exclusively under the WC model.

~~~~~~~~~~~~~~Best performers are bold. All experiments are repeated five times.

~~~~~~~~~~~~~~$^{**}p<0.05$; $^{***}p<0.001$.
\end{minipage}}
\end{tabular} }

\end{table*}

\begin{table*}[htp]
\centering
\caption{\jiadong{Comparision with Other Heuristics (Expected Spread) in IMIN-EB problem (LT model)}}\label{table:heuristics-eb-wc-lt2}
\resizebox{0.9\linewidth}{!}{
\begin{tabular}{c|lll|lll|lll|lll}
\toprule
\multirow{2}{*}{\centering b } & \multicolumn{3}{c|}{EmailCore (WC model)} & \multicolumn{3}{c|}{Facebook (WC model)} & \multicolumn{3}{c|}{Wiki-Vote (WC model)} & \multicolumn{3}{c}{EmailAll (WC model)} \\
& RA & OD & AG  & RA & OD & AG  & RA & OD & AG  & RA & OD & AG  \\ \midrule
20 & 150.71$^{***}$ & 140.09$^{***}$ & \bf{129.69} & 53.284$^{***}$ & 55.036$^{***}$ & \bf{28.192} & 22.971$^{***}$ & 22.527$^{***}$ & \bf{18.445} & 157.48$^{***}$ & 151.63$^{***}$ & \bf{116.66} \\
40 & 149.65$^{***}$ & 139.04$^{***}$ & \bf{114.99} & 50.290$^{***}$ & 53.601$^{***}$ & \bf{23.228} & 22.562$^{***}$ & 20.732$^{***}$ & \bf{15.708} & 155.84$^{***}$ & 151.33$^{***}$ & \bf{108.58} \\
60 & 139.11$^{***}$ & 136.82$^{***}$ & \bf{110.73} & 49.761$^{***}$ & 50.569$^{***}$ & \bf{21.943} & 20.840$^{***}$ & 20.499$^{***}$ & \bf{13.642} & 153.81$^{***}$ & 149.71$^{***}$ & \bf{88.054} \\
80 & 136.26$^{***}$ & 135.74$^{***}$ & \bf{96.138} & 49.173$^{***}$ & 50.370$^{***}$ & \bf{20.212} & 20.424$^{***}$ & 20.420$^{***}$ & \bf{11.509} & 142.38$^{***}$ & 143.98$^{***}$ & \bf{59.450} \\
100 & 133.45$^{***}$ & 132.10$^{***}$ & \bf{89.610} & 48.574$^{***}$ & 48.089$^{***}$ & \bf{19.159} & 20.415$^{***}$ & 20.379$^{***}$ & \bf{11.016}$^{***}$ & 138.71$^{***}$ & 135.55$^{***}$ & \bf{43.745} \\
\bottomrule

\toprule
\multirow{2}{*}{\centering b } & \multicolumn{3}{c|}{DBLP (WC model)} & \multicolumn{3}{c|}{Twitter (WC model)} & \multicolumn{3}{c|}{Stanford (WC model)} & \multicolumn{3}{c}{Youtube (WC model)} \\ 
& RA & OD & AG  & RA & OD & AG  & RA & OD & AG  & RA & OD & AG  \\  \midrule
20 & 16.479$^{***}$ & 16.249$^{***}$ & \bf{13.582} & 228.32$^{***}$ & 224.13$^{***}$ & \bf{211.76} & 32.991$^{***}$ & 32.914$^{***}$ & \bf{16.977} & 39.883$^{***}$ & 39.188$^{***}$ & \bf{28.027} \\
40 & 16.288$^{***}$ & 16.145$^{***}$ & \bf{13.353} & 225.67$^{***}$ & 218.33$^{***}$ & \bf{194.24} & 32.597$^{***}$ & 32.071$^{***}$ & \bf{14.491} & 39.217$^{***}$ & 39.182$^{***}$ & \bf{21.656} \\
60 & 15.766$^{***}$ & 15.928$^{***}$ & \bf{13.173} & 216.97$^{***}$ & 213.86$^{***}$ & \bf{191.15} & 32.446$^{***}$ & 32.068$^{***}$ & \bf{12.745} & 39.067$^{***}$ & 38.839$^{***}$ & \bf{16.904} \\
80 & 15.651$^{***}$ & 15.491$^{***}$ & \bf{12.024} & 209.50 $^{***}$& 208.04$^{***}$ & \bf{178.17} & 32.129$^{***}$ & 31.920$^{***}$ & \bf{11.588} & 38.591$^{***}$ & 37.354$^{***}$ & \bf{10.868} \\
100 & 15.305$^{***}$ & 14.824$^{***}$ & \bf{11.700} & 200.55$^{***}$ & 199.60$^{***}$ & \bf{176.21} & 30.733$^{***}$ & 30.664$^{***}$ & \bf{10.543} & 37.379$^{***}$& 36.745$^{***}$ & \bf{10} \\
\bottomrule
\multicolumn{13}{l}{\begin{minipage}{1.1\textwidth}  Notes. Since the LT model requires the probabilities of incoming edges for each node to be equal to or less than $1$, we choose to evaluate our algorithms exclusively under the WC model.

~~~~~~~~~~~~~~Best performers are bold. All experiments are repeated five times.

~~~~~~~~~~~~~~$^{***}p<0.001$.\end{minipage}}
\end{tabular} }
\end{table*}

\clearpage
\section*{Online Appendix 4: Figures}

\begin{figure}[!htbp]
\begin{subfigure}[h]{0.24\linewidth}
    \centering
    \includegraphics[width=\linewidth]{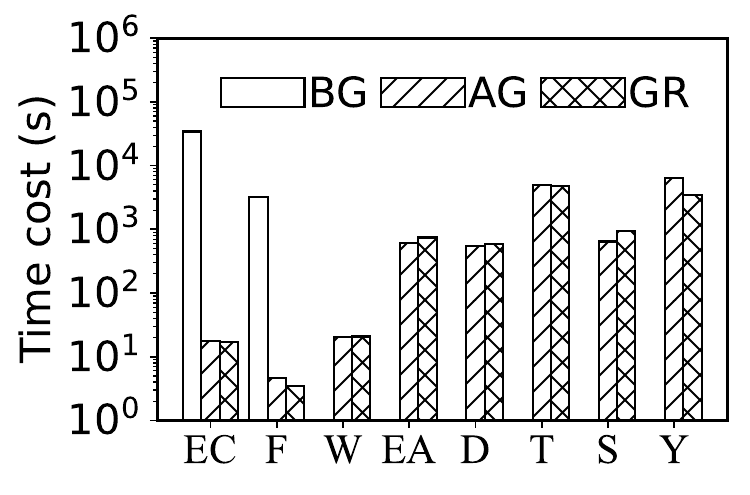}
    \caption{\jiadong{TR \& LT model in IMIN problem}}
    \label{fig:time-tr-lt}
    \end{subfigure}
    \hspace{1mm}
\begin{subfigure}[h]{0.24\linewidth}
    \centering
    \includegraphics[width=\linewidth]{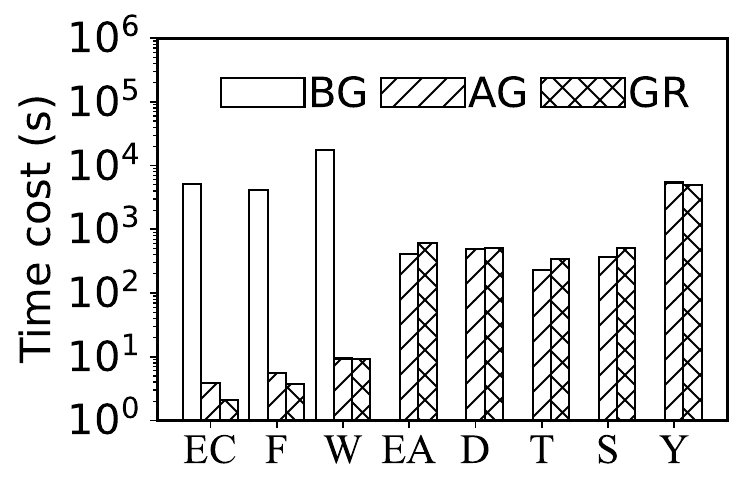}
    \caption{\jiadong{WC \& LT model in IMIN problem}}
    \label{fig:time-wc-lt}
    \end{subfigure}
    \hspace{1mm}
\begin{subfigure}[h]{0.24\linewidth}
    \centering
    \includegraphics[width=\linewidth]{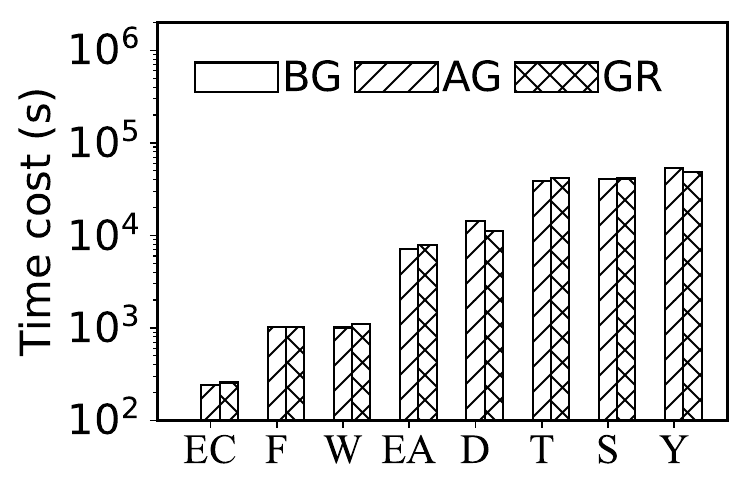}
    \caption{\jiadong{TR \& LT model in IMIN-EB problem}}
    \label{fig:time-tr-lt-eb}
    \end{subfigure}
    \hspace{1mm}
\begin{subfigure}[h]{0.24\linewidth}
    \centering
    \includegraphics[width=\linewidth]{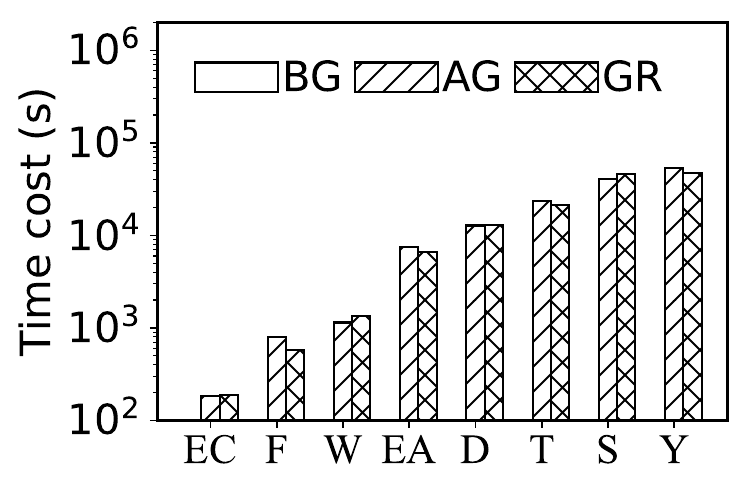}
    \caption{\jiadong{WC \& LT model in IMIN-EB problem}}
    \label{fig:time-wc=lt=eb}
    \end{subfigure}
    \caption{\jiadong{Time Cost of Different Algorithms}} \label{fig:time-imin-eb}
\end{figure}

\begin{figure}[!htbp]
\begin{minipage}{0.49\linewidth}
\begin{subfigure}[h]{0.48\linewidth}
    \centering
    \includegraphics[width=\linewidth]{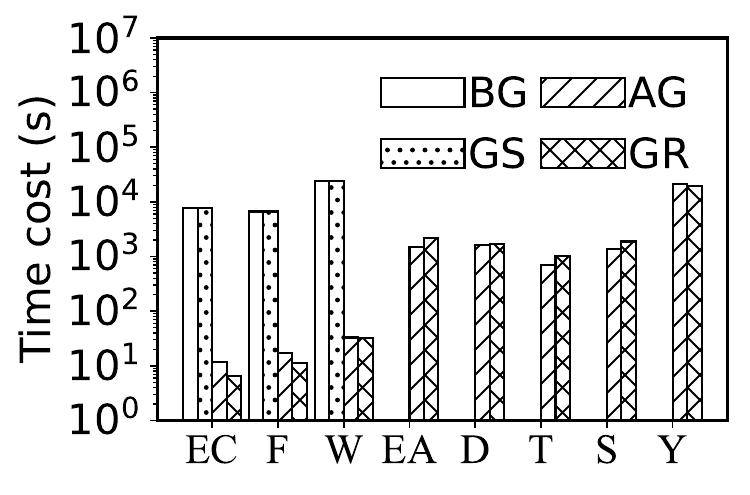}
    \caption{WC \& IC model in IMIN problem}
    \label{fig:time-wc}
    \end{subfigure}
\begin{subfigure}[h]{0.48\linewidth}
    \centering
    \includegraphics[width=\linewidth]{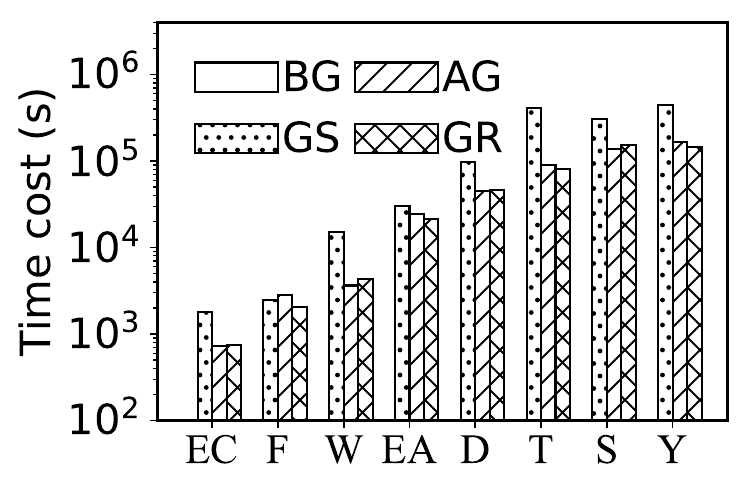}
    \caption{\jiadong{WC \& IC model in IMIN-EB problem}}
    \label{fig:time-wc-eb}
    \end{subfigure}
    \caption{Time Cost of Different Algorithms} \label{fig:time-imin-2}
    \end{minipage}
    \begin{minipage}{0.51\linewidth}
\begin{subfigure}[h]{0.49\linewidth}
\centering
    \includegraphics[width=1\columnwidth]{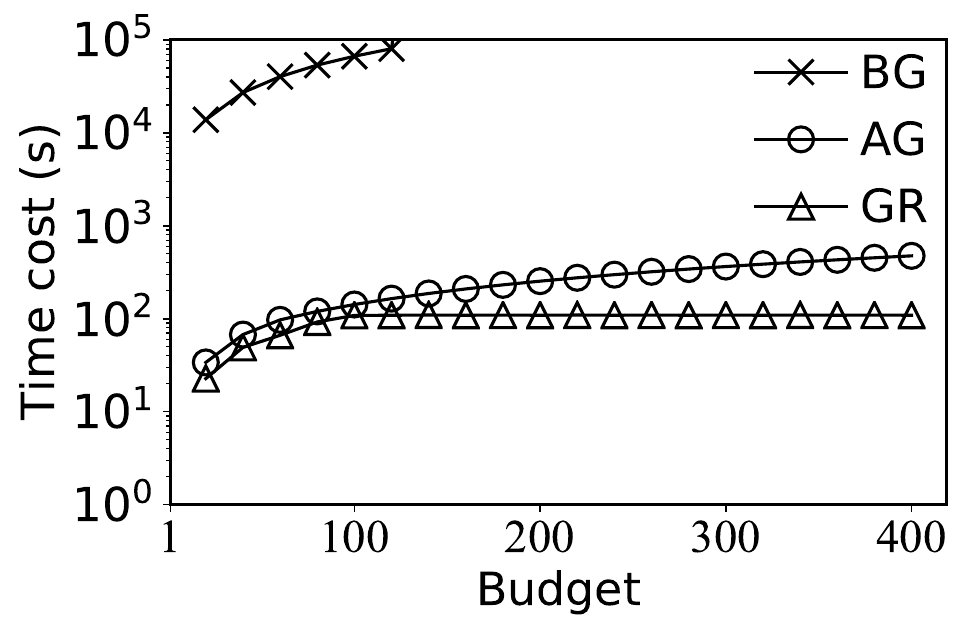}
    \caption{WC \& IC model in \texttt{Facebook} dataset}\label{fig:time-budget-b}
\end{subfigure}
\begin{subfigure}[h]{0.49\linewidth}
\centering
    \includegraphics[width=1\columnwidth]{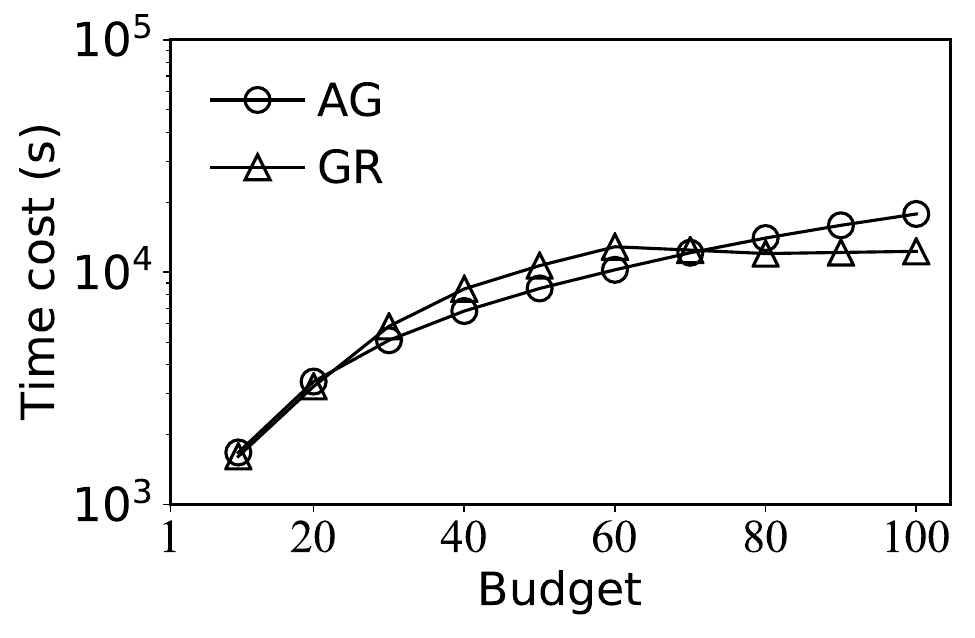}
    \caption{WC \& IC model in \texttt{DBLP} dataset}\label{fig:time-budget-d}
\end{subfigure}
\caption{Running Time v.s. Budget in the IMIN problem}
\label{fig:time-budget-2}
    \end{minipage}
\end{figure}

\clearpage
\section*{Online Appendix 5: Additional Results of Scalability Study}

\begin{figure}[!htbp]
\begin{subfigure}[h]{0.45\linewidth}
    \centering
    \includegraphics[width=\linewidth]{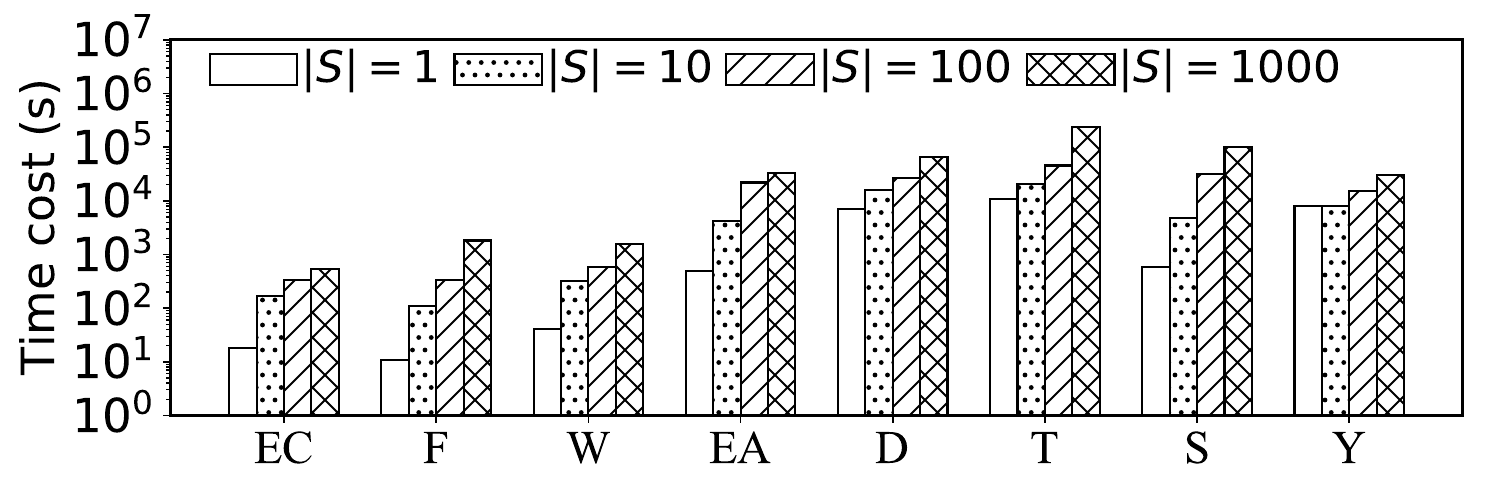}
    \caption{TR Model}
    \label{fig:sca-tr}
    \end{subfigure}
\begin{subfigure}[h]{0.45\linewidth}
    \centering
    \includegraphics[width=\linewidth]{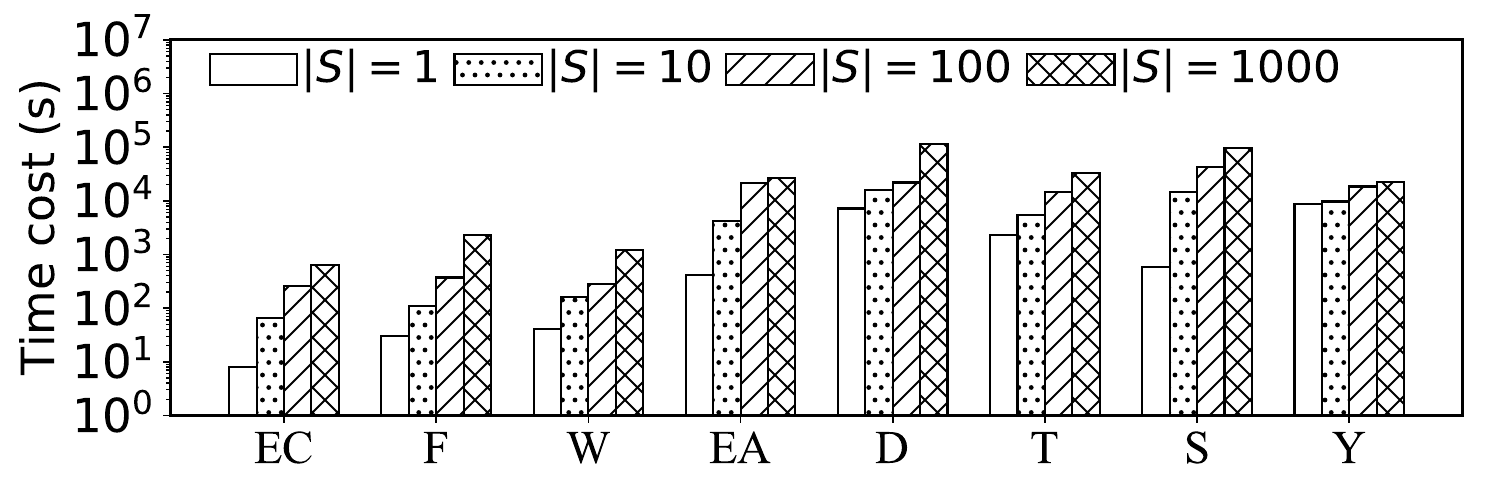}
    \caption{WC Model}
    \label{fig:sca-wc}
    \end{subfigure}
    \caption{Running Time v.s. Number of Seeds (IMIN problem under the IC model)}
\end{figure}

The scalability of our GR algorithm is assessed in Figures~\ref{fig:sca-tr} and~A5. We set the budget to $100$ and vary the number of seeds from $1,10,100$ to $1000$. We report the average time cost of the GR algorithm by executing it 5 times. We can see that as the number of seeds increases, so does the running duration. This is due to the fact that a greater number of seeds results in a greater influence spread (a larger size of sampled graphs), and the running duration of Algorithm~\ref{algo:blockes} is highly correlated with the size of sampled graphs. 
We can also observe that the proportional increase in running time is much less than the proportional increase in the number of seeds, proving that GR is scalable enough to manage scenarios with a large number of seeds.

\clearpage
\section*{Online Appendix 6: Additional Results of Study of $\theta$ Selection}

In this section, we provide the justification for selecting $\theta$ (i.e., the number of sampled graphs), which represents the number of sampled graphs used to choose the blocker in each round. 

In Figure~\ref{fig:es-theta} and Figure~\ref{fig:time-theta}, we vary $\theta$ from $10^3$, $10^4$ to $10^5$, and report the expected spread and running time of our GR algorithm. 
We evaluate on all datasets under the TR model by setting the blocker budget to $20$ and randomly selecting $10$ seed nodes. The y-axis is the decrease ratio of the expected spread\footnote{We show the decrease ratio because the absolute differences of expected spreads are quite small.} using three $\theta$ values in Figure~\ref{fig:es-theta} and the running time (s) in Figure~\ref{fig:time-theta}. The x-axis represents different networks whose abbreviations are listed in Table \ref{tab:dat} column (2).
From Figure ~\ref{fig:es-theta}, we observe that the largest decrease ratio from $\theta=10^3$ to $\theta=10^4$ is only $2.89\%$, and the largest decrease ratio from $\theta=10^4$ to $\theta=10^5$ is less than $0.1\%$.
Figure~\ref{fig:time-theta} shows the running time gradually increases when $\theta$ increases. 
Based on the obtained results, we selected a value of $\theta=10^4$ for all experiments, as it provided a favorable balance between time costs and accuracy. 



\begin{figure}[!htbp]
\begin{subfigure}[h]{0.47\linewidth}
    \centering
    \includegraphics[width=\linewidth]{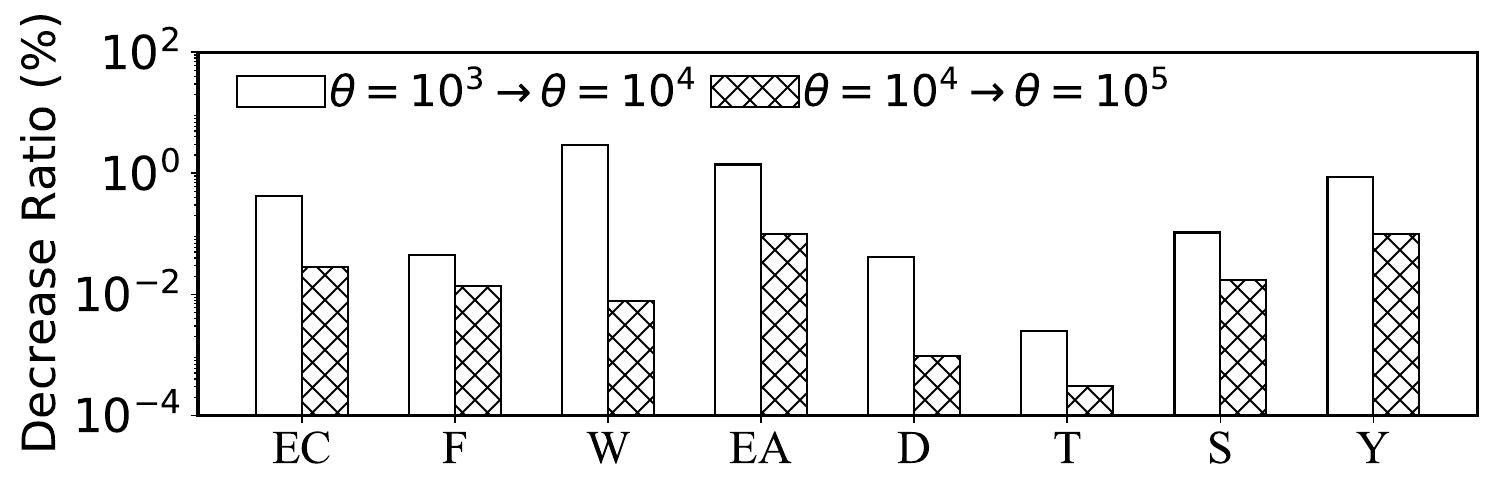}
    \caption{Expected Spread v.s. \#Sampled Graphs}
    \label{fig:es-theta}
    \end{subfigure}
\begin{subfigure}[h]{0.47\linewidth}
    \centering
    \includegraphics[width=\linewidth]{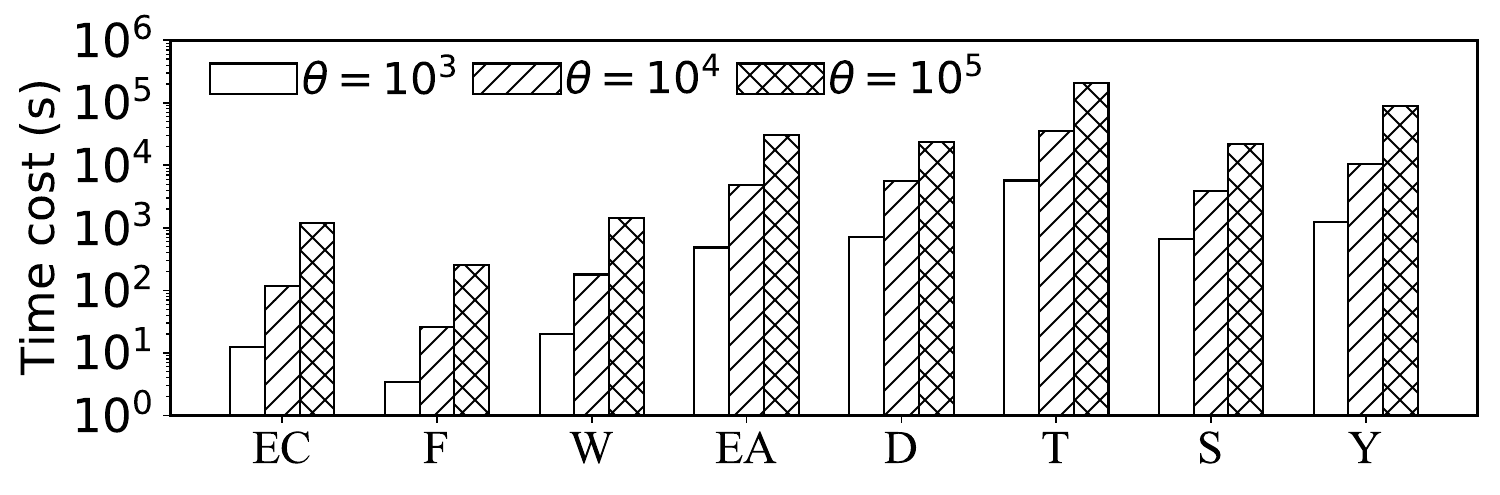}
    \caption{Running Time v.s. Number of Sampled Graphs}
    \label{fig:time-theta}
    \end{subfigure}
    \caption{Time Cost of Different Algorithms for the IMIN problem under the IC model} \label{fig:time}
\end{figure}